\documentclass[journal,12pt,draftclsnofoot,onecolumn]{IEEEtran} 

\usepackage{color}
\usepackage{balance}
\usepackage{graphicx}
\usepackage{enumerate}
\usepackage{subfigure}
\usepackage{algorithm}
\usepackage{algorithmic}
\usepackage{booktabs}
\usepackage{cite}
\usepackage{amssymb}
\usepackage{amsmath}
\newtheorem{proposition}{Proposition}
\newtheorem{theorem}{Theorem}

\newenvironment{proof}{{\indent \indent \it Proof:\quad}}{\hfill $\blacksquare$\par}

\begin{document}

\title{NOMA for Energy-Efficient LiFi-Enabled Bidirectional IoT Communication}
	
\author{
Chen Chen, \IEEEmembership{Member, IEEE,}
Shu Fu,
Xin Jian,
Min Liu, \\
Xiong Deng,
and Zhiguo Ding, \IEEEmembership{Fellow, IEEE}

\thanks{This work was supported by the National Natural Science Foundation of China under Grant 61901065 and Grant 61701054.
}

\thanks{C. Chen and S. Fu are with the School of Microelectronics and Communication Engineering, Chongqing University, Chongqing 400044, China. They are also with the State Key Laboratory of Integrated Services Networks, Xidian University, Xi'an, Shaanxi 710071, China. (e-mail: \{c.chen, shufu\}@cqu.edu.cn).}

\thanks{X. Jian and M. Liu are with the School of Microelectronics and Communication Engineering, Chongqing University, Chongqing 400044, China (e-mail: \{jianxin, liumin\}@cqu.edu.cn).}

\thanks{X. Deng is with the Department of Electrical Engineering, Eindhoven University of Technology (TU/e), Flux Building, 5600MB Eindhoven, Netherlands (e-mail: X.Deng@tue.nl).}

\thanks{Z. Ding is with the School of Electrical and Electronic Engineering, The University of Manchester, Manchester M13 9PL, U.K. (e-mail: zhiguo.ding@manchester.ac.uk).}
}

\markboth{}
{}

\maketitle

\begin{abstract}
In this paper, we consider a light fidelity (LiFi)-enabled bidirectional Internet of Things (IoT) communication system, where visible light and infrared light are used in the downlink and uplink, respectively. In order to improve the energy efficiency (EE) of the bidirectional LiFi-IoT system, non-orthogonal multiple access (NOMA) with a quality-of-service (QoS)-guaranteed optimal power allocation (OPA) strategy is applied to maximize the EE of the system. We derive a closed-form OPA set based on the identification of the optimal decoding orders in both downlink and uplink channels, which can enable low-complexity power allocation. Moreover, we propose an adaptive channel and QoS-based user pairing approach by jointly considering users' channel gains and QoS requirements. We further analyze the EE of the bidirectional LiFi-IoT system and the user outage probabilities (UOPs) of both downlink and uplink channels of the system. Extensive analytical and simulation results demonstrate the superiority of NOMA with OPA in comparison to orthogonal multiple access (OMA) and NOMA with typical channel-based power allocation strategies. It is also shown that the proposed adaptive channel and QoS-based user pairing approach greatly outperforms individual channel/QoS-based approaches, especially when users have diverse QoS requirements.
\end{abstract}

\begin{IEEEkeywords}
Non-orthogonal multiple access (NOMA), light fidelity (LiFi), Internet of Things (IoT), energy efficiency (EE), user outage probability (UOP).
\end{IEEEkeywords}

\section{Introduction}

{W}{ith} the explosive increase of smart devices in our everyday life, the Internet of Things (IoT) has been emerging as a promising solution to connect a large number of devices \cite{atzori2010internet}. The IoT paradigm contains a variety of devices such as electronic devices, mobile devices and industrial devices, and different devices can have different communication and computation capabilities and quality-of-service (QoS) requirements \cite{da2014internet}. As a key enabling technology of IoT, communication plays a vital role to connect all the smart devices supported in the IoT networks. Many radio frequency (RF)-based techniques have been considered for IoT communication such as RFID, ZigBee, Bluetooth, WiFi and 5G \cite{palattella2016internet}. Recently, Light Fidelity (LiFi) has been envisioned as a promising IoT communication technology, which provides many attractive features that the RF-based IoT networks might struggle to offer, including accurate device positioning, energy harvesting from light and inherent physical-layer security \cite{demirkol2019powering}. As a lightwave-based communication technology, LiFi aims to realize a fully networked bidirectional wireless communication system by exploiting visible light in the downlink and infrared light in the uplink \cite{haas2016lifi}. In particular, visible light-based LiFi downlink can be built upon the existing light emitting diode (LED) fixture which is widely deployed for general indoor lighting \cite{komine2004fundamental,burchardt2014vlc}.

\subsection{Related Work and Motivation}

Although LiFi reveals its potential for future IoT networks, the research of LiFi-enabled IoT is still at the early stage. In \cite{albraheem2018toward}, a LiFi-based hierarchical IoT architecture was proposed to analyze the collected data and build smart decisions. In \cite{diamantoulakis2017simultaneous} and \cite{sharma2018eh}, the energy harvesting issues of LiFi-IoT were investigated. Lately, a LiFi-IoT system vision was reported in \cite{demirkol2019powering}, where the conceptual architecture with four different types of motes was presented.

Considering that pervasive IoT is usually required to connect a huge number of IoT devices per unit area \cite{demirkol2019powering}, the LiFi access point (AP) of an optical attocell in LiFi-IoT networks should be able to support multiple IoT devices. Therefore, an efficient multiple access technique is of great significance to successfully implement LiFi-IoT in practical scenarios. So far, many multiple access techniques have been introduced for visible light-based downlink LiFi communication, i.e., visible light communication (VLC), which can be mainly divided into two categories: one is orthogonal multiple access (OMA) and the other is non-orthogonal multiple access (NOMA). For OMA schemes such as orthogonal frequency division multiple access (OFDMA) and time division multiple access (TDMA), users are allocated with different orthogonal time or frequency resources \cite{abdelhady2018downlink,sung2015orthogonal,hammouda2018resource}. 
Although OMA can eliminate mutual interference between users, its resource utilization is inefficient. In contrast, NOMA allows multiple users to simultaneously utilize all the time and frequency resources through power domain superposition coding (SPC) and successive interference cancellation (SIC) \cite{marshoud2016non}. Due to its efficient resource utilization, NOMA has been recognized as a promising multiple access technique for multi-user VLC systems \cite{yin2016performance,marshoud2017performance,lin2017experimental,chen2018performance,ren2018OE,chen2018flexible}. 

It has been shown that the performance gain of NOMA over OMA is mainly determined by the specific power allocation strategy adopted by NOMA \cite{ding2017survey}. For NOMA-based multi-user VLC systems, various channel-based power allocation strategies have been proposed such as gain ratio power allocation (GRPA) and normalized gain difference power allocation (NGDPA)  \cite{marshoud2016non,chen2018performance}. Moreover, to support a large number of users in the VLC system using NOMA, channel-based 
user pairing schemes have also been proposed to efficiently divide users into pairs \cite{almohimmah2018simple,janjua2020user}. Nevertheless, all the aforementioned works are only focused on the application of NOMA in visible light-based LiFi downlink. To apply NOMA in LiFi-enabled bidirectional IoT communication, the following two important issues should be taken into consideration:

\subsubsection{Energy Consumption}

In LiFi-enabled bidirectional IoT communication, energy consumption originates from two parts: one is the LiFi AP within each optical attocell and the other is the connected IoT devices. Particularly, most IoT devices rely on batteries and reducing the energy consumption to extend their battery life is a top concern \cite{da2014internet,palattella2016internet}. Therefore, it is of practical significance to design an energy-efficient multiple access technique for LiFi-IoT.

\subsubsection{Diverse Device QoS Requirements}

In practical LiFi-IoT systems, IoT devices can be divided into two categories: one includes low-speed devices such as environmental sensors and health monitors, and the other consists of high-speed devices such as multimedia-capable mobile phones \cite{da2014internet,teli2018optical}. As a result, it is necessary to take the diverse QoS requirements of IoT devices into account when designing a multiple access technique for LiFi-IoT.

\subsection{Main Contributions}

To address above-mentioned issues when applying NOMA, in this paper, we propose an energy-efficient NOMA technique for bidirectional LiFi-IoT communication. The main contributions of this work are summarized as follows:
\begin{itemize}
\item An energy-efficient NOMA technique is applied for the bidirectional LiFi-IoT system, which adopts a QoS-guaranteed optimal power allocation (OPA) strategy to maximize the energy efficiency (EE) of the system. The optimal decoding orders in both downlink and uplink channels are first identified and proved, and then the OPA set is obtained in a simple and closed form, which enables low-complexity power allocation.
\item Three user pairing approaches are studied in the NOMA-enabled bidirectional LiFi-IoT system, including channel-based, QoS-based, and adaptive channel and QoS-based approaches. The newly proposed adaptive channel and QoS-based user pairing approach dynamically selects from the channel-based and the QoS-based approaches to achieve a higher EE.
\item Both EE and user outage probability (UOP) are analyzed in the bidirectional LiFi-IoT system using different multiple access techniques. It is analytically proved that NOMA with OPA always achieves higher EE than OMA and NOMA with typical channel-based power allocation strategies. The calculations of downlink and uplink UOPs are also discussed.
\item Extensive analytical and simulation results are presented to evaluate the performance of different multiple access techniques in a typical bidirectional LiFi-IoT system. The obtained results demonstrate the superiority of NOMA adopting OPA with adaptive channel and QoS-based user pairing for energy-efficient bidirectional LiFi-IoT systems.
\end{itemize}

The remainder of this paper is organized as follows. Section II presents the system model. The principle of NOMA and its application for energy-efficient bidirectional LiFi-IoT communication are described in Section III. The EE and UOP of the bidirectional LiFi-IoT system using different multiple access techniques are analyzed in Section IV. Detailed analytical and simulation results are presented in Section V. Finally, Section VI concludes the paper.

\section{System Model}

We present the basic model of the bidirectional LiFi-IoT system in this section. The configuration of the bidirectional LiFi-IoT system is first introduced, and then the light propagation model and the noise model are further discussed.

\subsection{Bidirectional LiFi-IoT System Configuration}

\begin{figure}[!t]
\centering
{\includegraphics[width=.5\columnwidth]{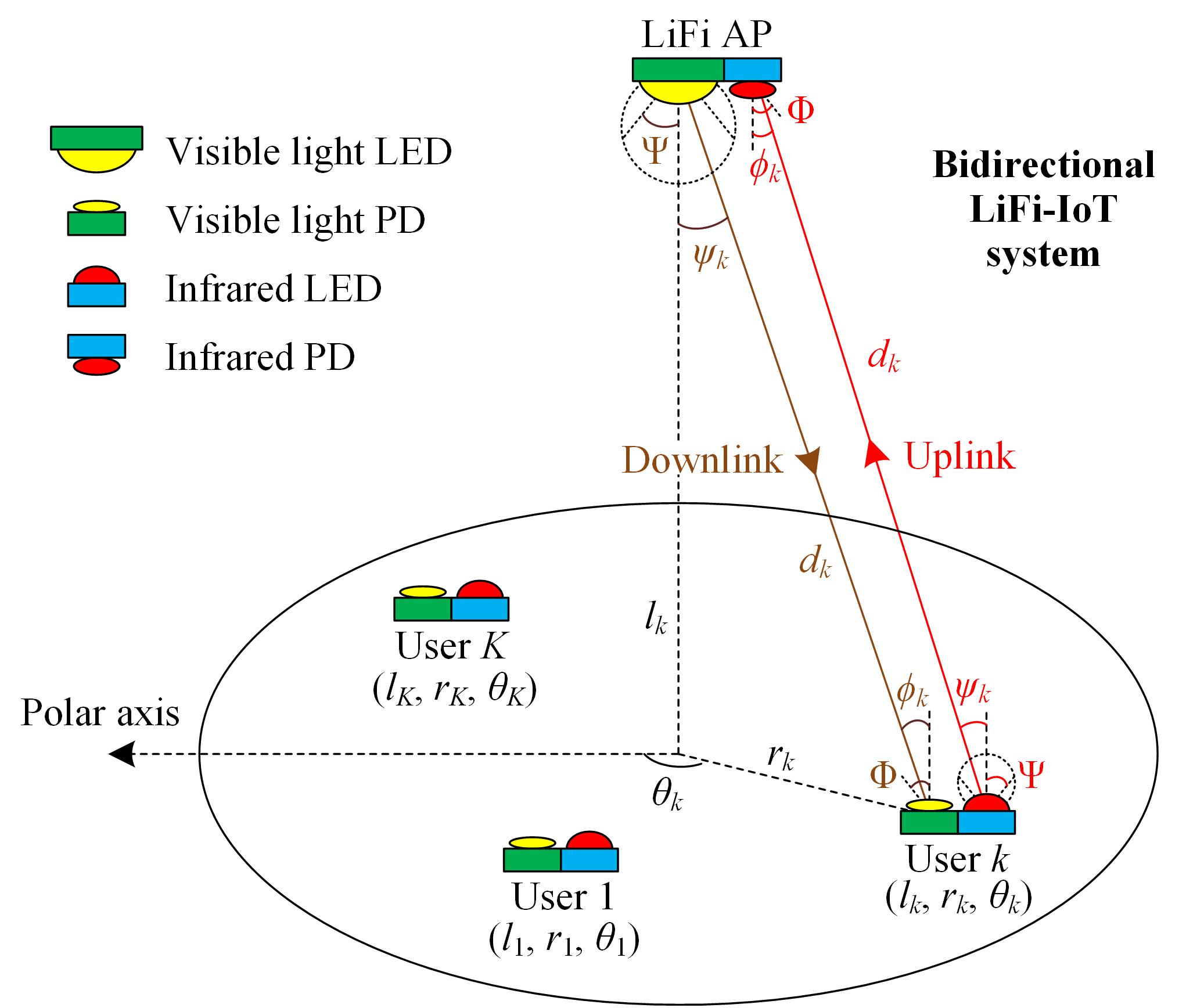}} 
\caption{Geometric configuration of the bidirectional LiFi-IoT system with one LiFi AP and $K$ users, i.e., IoT devices.}
\label{fig:System}
\end{figure}

In this work, we consider a bidirectional LiFi-IoT system, where visible light is utilized in the downlink for simultaneous illumination and communication, while infrared light is adopted for uplink communication. Fig.~\ref{fig:System} illustrates the geometric configuration of the bidirectional LiFi-IoT system with one LiFi AP and totally $K$ users, i.e., IoT devices. As we can see, the LiFi AP consists of  a visible light LED transmitter and an infrared light photodiode (PD) receiver, while each user is equipped with a visible light PD receiver and a infrared light LED transmitter. In the downlink, the visible light LED of the LiFi AP radiates white light to provide lighting within its coverage and broadcast downlink data to all the users at the same time. Each user detects the broadcasted data by using the equipped visible light PD. In the uplink, each user employs the equipped infrared light LED to transmit its uplink data and the LiFi AP utilizes the infrared light PD to collect the data from all the users. Therefore, the bidirectional communication between the LiFi AP and all the users in the LiFi-IoT system can be established.

Without loss of generality, we assume that both the visible and infrared light LEDs are point sources and they operate within their linear dynamic range. We further assume that the overall system has a flat frequency response. Moreover, we also assume that the visible light LED and the infrared light PD of the LiFi AP are oriented vertically downwards, while the visible light PD and the infrared light LED of each user are oriented vertically upwards. For simplicity, we assume that the visible/infrared light LEDs have the same semi-angle, while the visible/infrared light PDs have the same responsivity, active area and field-of-view (FOV). Under such assumptions, the models for visible light downlink and infrared light uplink channels become the same \cite{soltani2018bidirectional}.

\subsection{Light Propagation Model}

In practical LiFi-IoT systems, the visible light PD in the downlink and the infrared light PD in the uplink can receive both line-of-sight (LOS) and non-LOS components of the corresponding transmitted optical signals. Nevertheless, since the non-LOS component usually has much lower electrical power than that of the LOS component, it is reasonable to neglect the non-LOS component during most channel conditions \cite{zeng2009high}. For simplicity, we only consider the LOS component in the following channel model. 

For the visible and infrared light LEDs with a Lambertian emission pattern,  the LOS direct current (DC) channel gain between the LiFi AP and the $k$-th ($k = 1, 2, \cdots, K$) user for both downlink and uplink channels can be calculated by \cite{kahn1997wireless,komine2004fundamental}
\begin{equation}
\setlength{\abovedisplayskip}{12pt}
\setlength{\belowdisplayskip}{12pt}
h_k = 
\left\{
\begin{aligned}
\frac{(m+1) \rho A}{2\pi d_k^2} \mathrm{cos}^{m}(\psi_k) g_f g_l \mathrm{cos}(\phi_k),~0 \le \phi_k \le \Phi\\
0,~~~~~~~~~~~~~~~~~~~~~~~~~~~~~~~~~~~~~\phi_k > \Phi~~~~~\\
\end{aligned},
\right.	
\label{eqn:hk}
\end{equation}
where $m = -\mathrm{ln}2/\mathrm{ln}(\mathrm{cos}(\Psi))$ denotes the Lambertian emission order with $\Psi$ being the semi-angle of the visible/infrared light LED; $\rho$ and $A$ represent the responsivity and active area of the visible/infrared light PD, respectively; $d_k$ is the distance between the LiFi AP and the $k$-th user; $\psi_k$ and $\phi_k$ denote the corresponding emission angle and incident angle, respectively; $g_f$ and $g_l$ represent the gains of the optical filter and the optical lens, respectively. The gain of the optical lens can be calculated by $g_l = \frac{n^2}{\mathrm{sin}^2 \Phi}$, where $n$ is the refractive index of the optical lens and $\Phi$ is the half-angle field-of-view (FOV) of the PD.

Considering the fact that different users might have different heights in practical LiFi-IoT systems, the locations of users should be within a three-dimensional (3D) space. For better description of the 3D location of a user, the polar coordinate system is adopted as shown in Fig.~\ref{fig:System}. In the polar coordinate system, the 3D location of the $k$-th user can be represented by $(l_k, r_k, \theta_k)$, where $l_k$ and $r_k$ respectively denote its vertical and horizontal distances from the LiFi AP, and $\theta_k$ denotes its polar angle from the reference axis. 

As shown in Fig.~\ref{fig:System}, due to the assumption that the LiFi AP is oriented vertically downwards while each user is oriented vertically upwards, the emission angle and the incident angle corresponding to the LiFi AP and the $k$-th user become the same, i.e., $\psi_k = \phi_k$. Hence, $\psi_k$ and $\phi_k$ can be represented by $\psi_k = \phi_k = \mathrm{arctan}\left(\frac{r_k}{l_k}\right)$. Moreover, $d_k$ can be expressed by $d_k = \sqrt{l_k^2 + r_k^2}$. Therefore, $h_k$ can be rewritten as follows: 
\begin{equation}
\setlength{\abovedisplayskip}{12pt}
\setlength{\belowdisplayskip}{12pt}
h_k = 
\left\{
\begin{aligned}
\frac{\mathcal{C}}{l_k^2 + r_k^2} \mathrm{cos}^{m + 1}\left(\mathrm{arctan}\left(\frac{r_k}{l_k}\right)\right),~0 \le \frac{r_k}{l_k} \le \mathrm{tan} \Phi\\
0,~~~~~~~~~~~~~~~~~~~~~~~~~~~~~~~~~~~~~\frac{r_k}{l_k} > \mathrm{tan} \Phi~~~~~\\
\end{aligned}~,
\right.	
\label{eqn:hk2}
\end{equation}
where $\mathcal{C} = \frac{(m+1) \rho A g_f g_l}{2\pi}$. It can be found from (\ref{eqn:hk2}) that $h_k$ is dependent on both the vertical distance $l_k$ and the horizontal distance $r_k$, which is not affected by the polar angle $\theta_k$. 

\subsection{Noise Model}

The additive noises in both downlink and uplink channels consist of thermal and shot noises, which are generally modeled as real-valued zero-mean additive white Gaussian noises. For simplicity, it is assumed that the additive noises in both downlink and uplink channels have the same constant noise power spectral density (PSD) $N_0$. For a given signal bandwidth $B$, the noise powers can be obtained as $N_0 B$.

\section{NOMA for Bidirectional LiFi-IoT}

\subsection{Principle of NOMA}

\begin{figure}[!t]
\centering
{\includegraphics[width=.9\columnwidth]{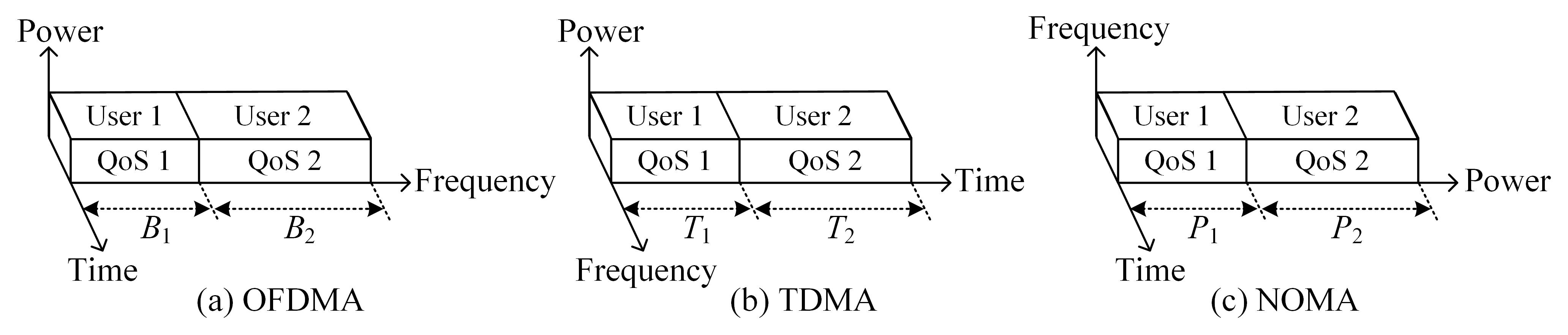}} 
\caption{Conceptual diagram of (a) OFDMA, (b) TDMA and (c) NOMA.}
\label{fig:Principle}
\end{figure}

Fig.~\ref{fig:Principle} shows the conceptual diagrams of OFDMA, TDMA and NOMA with two users, and the QoS requirements for user 1 and user 2 are represented by QoS 1 and QoS 2, respectively. For the case of OFDMA, the subcarriers are orthogonal to each other, and hence the QoS requirement of a specific user can be ensured by allocating it with a proper number of subcarriers, i.e., a proper bandwidth. As illustrated in Fig.~\ref{fig:Principle}(a), user 1 and user 2 are allocated with bandwidths $B_1$ and $B_2$ to satisfy their QoS requirements, respectively. Due to the orthogonality of all the subcarriers, there is no mutual interference between the transmitted data of two users. Nevertheless, interference-free multiple access is achieved by splitting the overall bandwidth into two parts. For the case of TDMA, as shown in Fig.~\ref{fig:Principle}(b), time slots $T_1$ and $T_2$ are allocated for user 1 and user 2 to meet their QoS requirements, respectively. Since time slots $T_1$ and $T_2$ are independent of each other, the transmitted data of two users are not mutually interfered. Hence, the mutual interference is eliminated by splitting the overall time resource into two parts. It can be concluded that the time or frequency resources must be split and shared so as to support multiple users without mutual interference for OMA schemes such as OFDMA and TDMA.

In contrast to OMA, NOMA allows both users to utilize all the time and frequency resources. It can be viewed from Fig.~\ref{fig:Principle}(c) that the transmitted data of user 1 and user 2 are superposed in the power domain and there inevitably exists mutual interference. To ensure their QoS requirements, user 1 and user 2 are allocated with powers $P_1$ and $P_2$, respectively. It has been well shown that the performance of NOMA is largely dependent on the adopted power allocation strategy \cite{ding2017survey}. In conventional NOMA-based systems, the power allocation strategy is  designed to maximize the sum rate of the system under a total transmit power constraint. However, when applying NOMA for energy-sensitive IoT applications, the power allocation strategy should be designed from the energy consumption perspective. Moreover, to support a large number of users, an efficient user pairing approach is also needed when implementing NOMA.

\subsection{NOMA for Downlink LiFi-IoT Using Visible Light}

\begin{figure}[!t]
\centering
{\includegraphics[width=.99\columnwidth]{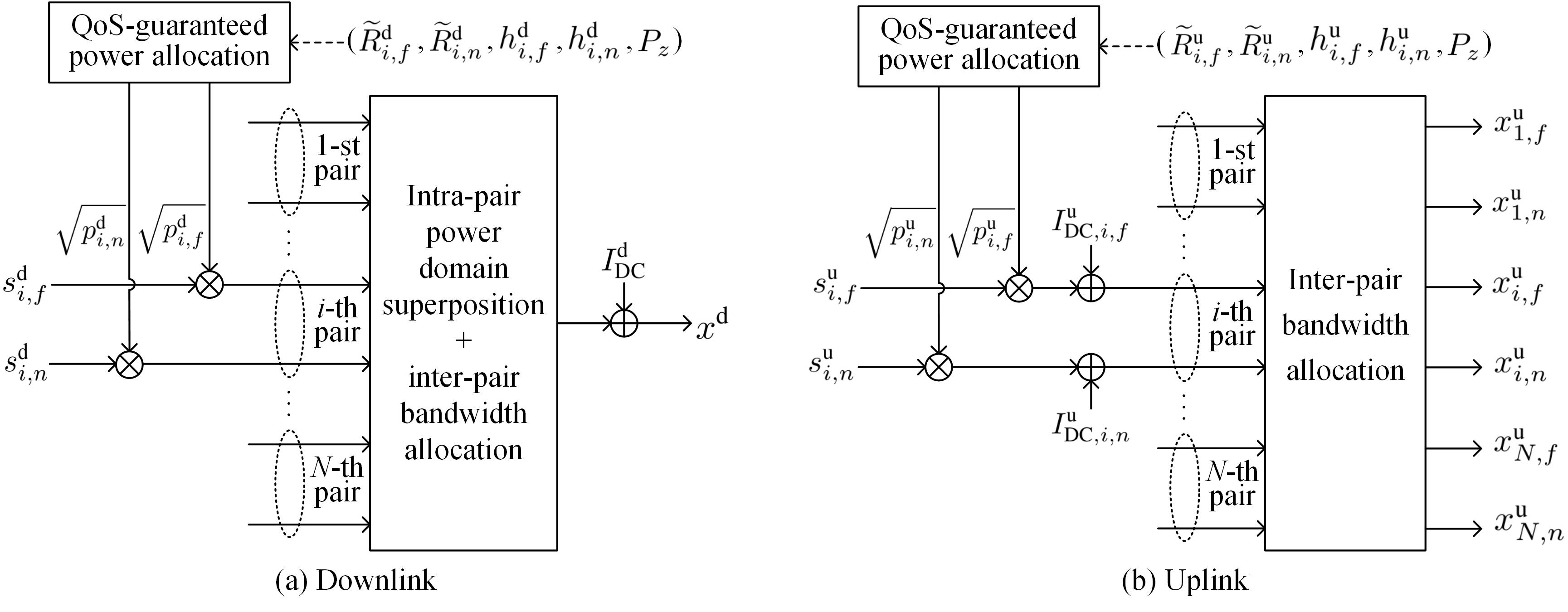}} 
\caption{Schematic diagram of NOMA-enabled LiFi-IoT (a) downlink and (b) uplink.}
\label{fig:Downlinkuplink}
\end{figure}

In this subsection, NOMA is introduced for downlink LiFi-IoT communication using visible light. Without loss of generality, we assume that the bidirectional LiFi-IoT system serves $K = 2 N$ users\footnote{Although a even number of users is considered here, NOMA is generally applicable to an arbitrary number of users. For an odd number of users, they are first sorted and then divided into pairs, and the remaining unpaired user can be allocated with separate power and bandwidth resources \cite{janjua2020user}.}, which are divided into $N$ user pairs. Fig. \ref{fig:Downlinkuplink}(a) shows the schematic diagram of NOMA-enabled LiFi-IoT downlink. Let $s_{i,f}^{\textrm{d}}$ and $s_{i,n}^{\textrm{d}}$ denote the modulated message signals intended for the far and near users in the $i$-th user pair, respectively. For the two users within each user pair, intra-pair power domain superposition is performed. Hence, the superposed electrical signal of the $i$-th user pair can be expressed by
\begin{equation}
\setlength{\abovedisplayskip}{12pt}
\setlength{\belowdisplayskip}{12pt}
x^{\textrm{d}}_i = \sqrt{p_{i,f}^{\textrm{d}}} s_{i,f}^{\textrm{d}} + \sqrt{p_{i,n}^{\textrm{d}}} s_{i,n}^{\textrm{d}},
\label{eqn:xdi}
\end{equation}
where $p_{i,f}^{\textrm{d}}$ and $p_{i,n}^{\textrm{d}}$ are the downlink electrical transmit powers allocated to the far and near users in the $i$-th user pair, respectively. After intra-pair power domain superposition, inter-pair bandwidth allocation is carried out and a DC bias current $I_{\textrm{DC}}^{\textrm{d}}$ is added to the resultant signal so as to simultaneously ensure the non-negativity of the driving signal of the visible light LED in the LiFi AP and guarantee sufficient and stable illumination. The total downlink electrical transmit power allocated to all $N$ pairs of users is obtained by $P_{\textrm{elec}}^{\textrm{d}} = \sum_{i = 1}^{N} p_{i,f}^{\textrm{d}} + p_{i,n}^{\textrm{d}}$.

After removing the DC term, the received downlink signals of the far and near users in the $i$-th user pair can be given by
\begin{equation}
\setlength{\abovedisplayskip}{12pt}
\setlength{\belowdisplayskip}{12pt}
  \left\{
  \begin{aligned}
  y_{i,f}^{\textrm{d}} = h_{i,f}^{\textrm{d}} (\sqrt{p_{i,f}^{\textrm{d}}} s_{i,f}^{\textrm{d}} + \sqrt{p_{i,n}^{\textrm{d}}} s_{i,n}^{\textrm{d}}) + z_{i,f}^{\textrm{d}}\\
  y_{i,n}^{\textrm{d}} = h_{i,n}^{\textrm{d}} (\sqrt{p_{i,f}^{\textrm{d}}} s_{i,f}^{\textrm{d}} + \sqrt{p_{i,n}^{\textrm{d}}} s_{i,n}^{\textrm{d}} ) + z_{i,n}^{\textrm{d}}\\
  \end{aligned}~,
  \right.
\label{eqn:yd}
\end{equation}
where $h_{i,f}^{\textrm{d}}$ and $h_{i,n}^{\textrm{d}}$ denote the downlink channel gains of the far and near users in the $i$-th user pair, respectively, and $h_{i,f}^{\textrm{d}} \le h_{i,n}^{\textrm{d}}$; $z_{i,f}^{\textrm{d}}$ and $z_{i,n}^{\textrm{d}}$ are the corresponding additive noises.

To decode the intended message signals for the far and near users in the $i$-th user pair, the decoding order should be first obtained. Here, we assume that the two users are sorted as a high priority user and a low priority user based on a specific sorting criteria, i.e., the decoding order can be generally given by $\mathbb{O}_{i,\textrm{high}}^{\textrm{d}} \ge \mathbb{O}_{i,\textrm{low}}^{\textrm{d}}$. The determination of the optimal decoding order will be discussed in Section III.D. Moreover, different users might have different QoS requirements in practical LiFi-IoT systems. Generally, we can define the QoS requirement of a specific user as its required achievable rate per bandwidth, i.e., spectral efficiency \cite{yin2016performance}. Let $\widetilde{R}_{i,\textrm{high}}^{\textrm{d}}$ and $\widetilde{R}_{i,\textrm{low}}^{\textrm{d}}$ denote the rate requirements of the high and low priority users in the $i$-th user pair in the downlink, respectively. The corresponding power requirements for both users in the $i$-th user pair in the downlink to meet their QoS requirements are given by the following theorem.

\begin{theorem}
For the high and low priority users in the $i$-th user pair with arbitrary QoS requirements in the downlink of the bidirectional LiFi-IoT system, the power requirements to satisfy their QoS requirements are given by
\begin{equation}
\setlength{\abovedisplayskip}{12pt}
\setlength{\belowdisplayskip}{12pt}
  \left\{
  \begin{aligned}
   p_{i,\textrm{high}}^{\textrm{d}} \ge 2^{2 \widetilde{R}_{i,\textrm{high}}^{\textrm{d}}} \left(2^{2 \widetilde{R}_{i,\textrm{low}}^{\textrm{d}}} \frac{P_z}{(h_{i,\textrm{low}}^{\textrm{d}})^2} + \frac{P_z}{(h_{i,\textrm{min}}^{\textrm{d}})^2 } \right)\\
   p_{i,\textrm{low}}^{\textrm{d}} \ge 2^{2 \widetilde{R}_{i,\textrm{low}}^{\textrm{d}}} \frac{P_z}{(h_{i,\textrm{low}}^{\textrm{d}})^2}~~~~~~~~~~~~~~~~~~~~~~~~~~~\\
  \end{aligned}~,
  \right.
\label{eqn:pd}
\end{equation}
where $h_{i,\textrm{min}}^{\textrm{d}} = \textrm{min} \{h_{i,\textrm{high}}^{\textrm{d}}, h_{i,\textrm{low}}^{\textrm{d}}\}$.
\end{theorem}

\begin{proof} Please refer to the appendix.
\end{proof}

Theorem 1 demonstrates that the QoS requirements of the high and low priority users in the $i$-th user pair in the downlink can be guaranteed under the impact of mutual interference by allocating them with proper powers.

\subsection{NOMA for Uplink LiFi-IoT Using Infrared Light}

Besides the downlink LiFi-IoT communication as discussed above, NOMA can also be applied for uplink LiFi-IoT communication using infrared light. The schematic diagram of NOMA-enabled LiFi-IoT uplink is plotted in Fig. \ref{fig:Downlinkuplink}(b). Let $s_{i,f}^{\textrm{u}}$ and $s_{i,n}^{\textrm{u}}$ be the modulated uplink message signals intended for the LiFi AP from the far and near users in the $i$-th user pair, respectively. The electrical signals to be transmitted by the infrared light LEDs of the far and near users in the $i$-th user pair can be expressed by
\begin{equation}
\setlength{\abovedisplayskip}{12pt}
\setlength{\belowdisplayskip}{12pt}
  \left\{
  \begin{aligned}
  x_{i,f}^{\textrm{u}} = \sqrt{p_{i,f}^{\textrm{u}}} s_{i,f}^{\textrm{u}} + I_{\textrm{DC},i,f}^{\textrm{u}}\\
  x_{i,n}^{\textrm{u}} = \sqrt{p_{i,n}^{\textrm{u}}} s_{i,n}^{\textrm{u}} + I_{\textrm{DC},i,n}^{\textrm{u}}\\
  \end{aligned}~,
  \right.
\label{eqn:xu}
\end{equation}
where $p_{i,f}^{\textrm{u}}$ and $p_{i,n}^{\textrm{u}}$ are the uplink electrical transmit powers of the far and near users in the $i$-th user pair, respectively; $I_{\textrm{DC},i,f}^{\textrm{u}}$ and $I_{\textrm{DC},i,n}^{\textrm{u}}$ are the DC bias currents added to guarantee the non-negativity of the driving signals of the infrared LEDs. The total uplink electrical transmit power of all $N$ pairs of users is given by $P_{\textrm{elec}}^{\textrm{u}} = \sum_{i = 1}^{N} p_{i,f}^{\textrm{u}} + p_{i,n}^{\textrm{u}}$.

At the LiFi AP, the received uplink signal of the $i$-th user pair after removing the DC term can be expressed by
\begin{equation}
\setlength{\abovedisplayskip}{12pt}
\setlength{\belowdisplayskip}{12pt}
y_i^{\textrm{u}} = h_{i,f}^{\textrm{u}} \sqrt{p_{i,f}^{\textrm{u}}} s_{i,f}^{\textrm{u}} + h_{i,n}^{\textrm{u}} \sqrt{p_{i,n}^{\textrm{u}}} s_{i,n}^{\textrm{u}} + z_{i}^{\textrm{u}},
\label{eqn:yui}
\end{equation}
where $h_{i,f}^{\textrm{u}}$ and $h_{i,n}^{\textrm{u}}$ are the uplink channel gains of the far and near users in the $i$-th user pair, respectively, and $h_{i,f}^{\textrm{u}} \le h_{i,n}^{\textrm{u}}$; $z_{i}^{\textrm{u}}$ is the corresponding additive noise.

Similarly, we assume that the far and near users in the $i$-th user pair in the uplink are sorted as a high priority user and a low priority user with the decoding order $\mathbb{O}_{i,\textrm{high}}^{\textrm{u}} \ge \mathbb{O}_{i,\textrm{low}}^{\textrm{u}}$. In addition, the QoS requirements of the high and low priority users in the $i$-th user pair are given by $\widetilde{R}_{i,\textrm{high}}^{\textrm{u}}$ and $\widetilde{R}_{i,\textrm{low}}^{\textrm{u}}$, respectively. The following theorem gives the power requirements for the high and low priority users in the $i$-th user pair in the uplink to meet their QoS requirements.

\begin{theorem}
For the high and low priority users in the $i$-th user pair with arbitrary QoS requirements in the uplink of the bidirectional LiFi-IoT system, the required powers to meet their QoS requirements are given by
\begin{equation}
\setlength{\abovedisplayskip}{12pt}
\setlength{\belowdisplayskip}{12pt}
  \left\{
  \begin{aligned}
   p_{i,\textrm{high}}^{\textrm{u}} \ge 2^{2 \widetilde{R}_{i,\textrm{high}}^{\textrm{u}}} \frac{(1 + 2^{2 \widetilde{R}_{i,\textrm{low}}^{\textrm{u}}}) P_z}{(h_{i,\textrm{high}}^{\textrm{u}})^2}\\
   p_{i,\textrm{low}}^{\textrm{u}} \ge 2^{2 \widetilde{R}_{i,\textrm{low}}^{\textrm{u}}} \frac{P_z}{(h_{i,\textrm{low}}^{\textrm{u}})^2}~~~~~~~~~\\
  \end{aligned}~.
  \right.
\label{eqn:pu}
\end{equation}
\end{theorem}

\begin{proof} Please refer to the appendix.
\end{proof}

It is demonstrated by Theorem 2 that, by adopted a proper power allocation strategy, the QoS requirements of the mutually interfered high and low priority users in the $i$-th user pair in the uplink can be successfully guaranteed.

\subsection{QoS-Guaranteed Optimal Power Allocation (OPA)}

For the efficient implementation of NOMA in bidirectional LiFi-IoT systems, a QoS-guaranteed OPA strategy is derived in this subsection. In practical LiFi-IoT systems, each user has its own QoS requirements, i.e., rate requirements, in both downlink and uplink channels. As a result, a bidirectional LiFi-IoT system is considered working properly, as long as the QoS requirements of all the users are satisfied. This is quite different from conventional NOMA-based systems, since their common goal is to maximize the achievable sum rate of all the users under a total transmit power constraint \cite{zhang2017user}. As energy consumption is a very important factor that needs to be considered when designing an IoT system, it is of practical significance to reduce the energy consumption of the system without compromising its working performance. From the energy consumption perspective, the goal to design NOMA-enabled bidirectional LiFi-IoT systems is to maximize the system EE via a QoS-guaranteed OPA strategy. Generally, the EE ($\eta$) with unit bits/J/Hz can be defined as the ratio of the achievable sum rate ($R$) of the system to the total electrical power consumption ($P_{\textrm{elec}}$):
\begin{equation}
\setlength{\abovedisplayskip}{12pt}
\setlength{\belowdisplayskip}{12pt}
\eta = \frac{R}{P_{\textrm{elec}}}, 
\label{eqn:EE}
\end{equation}
where $P_{\textrm{elec}} = P_{\textrm{elec}}^{\textrm{d}} + P_{\textrm{elec}}^{\textrm{u}}$ is the total power consumption of both downlink and uplink channels. 

For the NOMA-enabled bidirectional LiFi-IoT system with $N$ pairs of users, the target sum rate of the system can be calculated by
\begin{equation}
\setlength{\abovedisplayskip}{12pt}
\setlength{\belowdisplayskip}{12pt}
R = \sum_{i = 1}^{N} \widetilde{R}_{i,f}^{\textrm{d}} + \widetilde{R}_{i,n}^{\textrm{d}} + \widetilde{R}_{i,f}^{\textrm{u}} + \widetilde{R}_{i,n}^{\textrm{u}},
\label{eqn:R}
\end{equation}
where $\widetilde{R}_{i,f}^{\textrm{d}}$ and $\widetilde{R}_{i,n}^{\textrm{d}}$ denote the rate requirements of the far and near users in the $i$-th user pair in the downlink, while $\widetilde{R}_{i,f}^{\textrm{u}}$ and $\widetilde{R}_{i,n}^{\textrm{u}}$ denote the rate requirements of the far and near users in the $i$-th user pair in the uplink. 

\subsubsection{Optimal Decoding Order}
To obtain the OPA strategy, the optimal decoding orders for the far and near users in the $i$-th user pair in both downlink and uplink channels should be first identified. Based on the derived power requirements in (\ref{eqn:pd}) and (\ref{eqn:pu}), the optimal decoding orders are given by the following proposition.

\begin{proposition}
The optimal decoding orders for the far and near users in the $i$-th user pair in the downlink and uplink of the bidirectional LiFi-IoT system are given by $\mathbb{O}_{i,f}^{\textrm{d}} \ge \mathbb{O}_{i,n}^{\textrm{d}}$ and $\mathbb{O}_{i,f}^{\textrm{u}} < \mathbb{O}_{i,n}^{\textrm{u}}$, respectively.
\end{proposition}

\begin{proof} Please refer to the appendix.
\end{proof}

According to Proposition 1, the power requirements for the far and near users in the $i$-th user pair in both downlink and uplink channels of the bidirectional LiFi-IoT system are obtained by
\begin{equation}
\setlength{\abovedisplayskip}{12pt}
\setlength{\belowdisplayskip}{12pt}
  \left\{
  \begin{aligned}
   p_{i,f}^{\textrm{d}} \ge 2^{2 \widetilde{R}_{i,f}^{\textrm{d}}} \left( 2^{2 \widetilde{R}_{i,n}^{\textrm{d}}} \frac{P_z}{(h_{i,n}^{\textrm{d}})^2} + \frac{P_z}{(h_{i,f}^{\textrm{d}})^2 } \right)\\
   p_{i,n}^{\textrm{d}} \ge 2^{2 \widetilde{R}_{i,n}^{\textrm{d}}} \frac{P_z}{(h_{i,n}^{\textrm{d}})^2}~~~~~~~~~~~~~~~~~~~~~~~\\
p_{i,f}^{\textrm{u}} \ge 2^{2 \widetilde{R}_{i,f}^{\textrm{u}}} \frac{P_z}{(h_{i,f}^{\textrm{u}})^2}~~~~~~~~~~~~~~~~~~~~~~~\\
p_{i,n}^{\textrm{u}} \ge 2^{2 \widetilde{R}_{i,n}^{\textrm{u}}} \frac{(1 + 2^{2 \widetilde{R}_{i,f}^{\textrm{u}}}) P_z}{(h_{i,n}^{\textrm{u}})^2}~~~~~~~~~~~~~~\\
  \end{aligned}~.
  \right.
\label{eqn:pdu}
\end{equation}

\subsubsection{Problem Formulation}
Let $\mathcal{P}_i = \{p_{i,f}^{\textrm{d}}, p_{i,n}^{\textrm{d}}, p_{i,f}^{\textrm{u}}, p_{i,n}^{\textrm{u}}\}$ denote the power allocation set for the far and near users in the $i$-th user pair in both downlink and uplink channels. To obtain a QoS-guaranteed OPA strategy, i.e., optimal $\mathcal{P}_i$ with $i = 1, 2, \cdots, N$, the EE maximization problem of the NOMA-enabled bidirectional LiFi-IoT system is formulated as
\begin{equation}
\setlength{\abovedisplayskip}{12pt}
\setlength{\belowdisplayskip}{12pt}
\begin{split}
& \max_{\{\mathcal{P}_1 , \cdots, \mathcal{P}_N\}} \eta \\
& \textrm{s.t.}~~ \textrm{C1:}~(\ref{eqn:pdu})\\
& ~~~~~\textrm{C2:}~P_{\textrm{elec}}^{\textrm{d}} \le P_{\textrm{max}}^{\textrm{d}}\\
& ~~~~~\textrm{C3:}~p_{i,f}^{\textrm{u}} \le p_{\textrm{max}}^{\textrm{u}},~i \in \{1, 2, \cdots, N\}\\
& ~~~~~\textrm{C4:}~p_{i,n}^{\textrm{u}} \le p_{\textrm{max}}^{\textrm{u}},~i \in \{1, 2, \cdots, N\},
\end{split}
\label{eqn:EE2}
\end{equation}
where constraint ``C1'' is to guarantee the power requirements of all the users so as to meet their QoS requirements, constraint ``C2'' is that the total downlink electrical transmit power of the LiFi AP should not exceed its maximum value $P_{\textrm{max}}^{\textrm{d}}$, and constraints ``C3'' and ``C4'' are imposed to ensure that the uplink electrical transmit power does not exceed the maximum value $p_{\textrm{max}}^{\textrm{u}}$.

Considering the fact that each user in the bidirectional LiFi-IoT system normally has its fixed downlink and uplink QoS requirements during a period of time, the target sum rate $R$ given in (\ref{eqn:R}) can be viewed as a fixed value. Therefore, the EE maximization problem in (\ref{eqn:EE2}) can be transformed into a power minimization problem given as follows:
\begin{equation}
\setlength{\abovedisplayskip}{12pt}
\setlength{\belowdisplayskip}{12pt}
\begin{split}
& \min_{\{\mathcal{P}_1 , \cdots, \mathcal{P}_N\}}{P_{\textrm{elec}}}\\
& \textrm{s.t.}~~ \textrm{C1:}~(\ref{eqn:pdu})\\
& ~~~~~\textrm{C2:}~P_{\textrm{elec}}^{\textrm{d}} \le P_{\textrm{max}}^{\textrm{d}}\\
& ~~~~~\textrm{C3:}~p_{i,f}^{\textrm{u}} \le p_{\textrm{max}}^{\textrm{u}},~i \in \{1, 2, \cdots, N\}\\
& ~~~~~\textrm{C4:}~p_{i,n}^{\textrm{u}} \le p_{\textrm{max}}^{\textrm{u}},~i \in \{1, 2, \cdots, N\}.
\end{split}
\label{eqn:EE3}
\end{equation}

\subsubsection{Optimal Solution}
According to (\ref{eqn:pdu}), given the channel gains, QoS requirements and noise power, the power requirements for the far and near users in the $i$-th user pair in both downlink and uplink channels are independent from each other. Hence, the optimal solution for the power minimization problem is that the far and near users in the $i$-th user pair in both the downlink and uplink are allocated with minimum powers to satisfy their QoS requirements. Therefore, the closed-form OPA set 
$\mathcal{P}_i^{\textrm{opt}} = \{p_{i,f}^{\textrm{d,opt}}, p_{i,n}^{\textrm{d,opt}}, p_{i,f}^{\textrm{u,opt}}, p_{i,n}^{\textrm{u,opt}}\}$ can be obtained by
\begin{equation}
\setlength{\abovedisplayskip}{12pt}
\setlength{\belowdisplayskip}{12pt}
  \left\{
  \begin{aligned}
   p_{i,f}^{\textrm{d,opt}} = 2^{2 \widetilde{R}_{i,f}^{\textrm{d}}} \left( 2^{2 \widetilde{R}_{i,n}^{\textrm{d}}} \frac{P_z}{(h_{i,n}^{\textrm{d}})^2} + \frac{P_z}{(h_{i,f}^{\textrm{d}})^2 } \right)\\
   p_{i,n}^{\textrm{d,opt}} = 2^{2 \widetilde{R}_{i,n}^{\textrm{d}}} \frac{P_z}{(h_{i,n}^{\textrm{d}})^2}~~~~~~~~~~~~~~~~~~~~~~~\\
p_{i,f}^{\textrm{u,opt}} = 2^{2 \widetilde{R}_{i,f}^{\textrm{u}}} \frac{P_z}{(h_{i,f}^{\textrm{u}})^2}~~~~~~~~~~~~~~~~~~~~~~~\\
p_{i,n}^{\textrm{u,opt}} = 2^{2 \widetilde{R}_{i,n}^{\textrm{u}}} \frac{(1 + 2^{2 \widetilde{R}_{i,f}^{\textrm{u}}}) P_z}{(h_{i,n}^{\textrm{u}})^2}~~~~~~~~~~~~~~\\
  \end{aligned}~.
  \right.
\label{eqn:pdu2}
\end{equation}

Using (\ref{eqn:pdu2}), the minimum total electrical transmit power of all the $N$ pairs of users in both downlink and uplink channels of bidirectional LiFi-IoT system is given by
\begin{equation}
\setlength{\abovedisplayskip}{12pt}
\setlength{\belowdisplayskip}{12pt}
P_{\textrm{elec,min}}^{\textrm{NOMA}} = \sum_{i = 1}^{N} 2^{2 \widetilde{R}_{i,f}^{\textrm{d}}} \left( 2^{2 \widetilde{R}_{i,n}^{\textrm{d}}} \frac{P_z}{(h_{i,n}^{\textrm{d}})^2} + \frac{P_z}{(h_{i,f}^{\textrm{d}})^2 } \right) + 2^{2 \widetilde{R}_{i,n}^{\textrm{d}}} \frac{P_z}{(h_{i,n}^{\textrm{d}})^2} + 2^{2 \widetilde{R}_{i,f}^{\textrm{u}}} \frac{P_z}{(h_{i,f}^{\textrm{u}})^2} + 2^{2 \widetilde{R}_{i,n}^{\textrm{u}}} \frac{(1 + 2^{2 \widetilde{R}_{i,f}^{\textrm{u}}}) P_z}{(h_{i,n}^{\textrm{u}})^2}.
\label{eqn:Pmin}
\end{equation}

It should be noticed that the QoS-guaranteed OPA strategy is obtained based on the assumption that the $2 N$ users are divided into $N$ user pairs. Therefore, efficient user pairing should be first performed before executing the QoS-guaranteed OPA strategy within each user pair.

\subsection{User Pairing} 

By selecting a pair of users to perform NOMA, the computational complexity of the bidirectional LiFi-IoT system can be substantially reduced, which results in a hybrid multiple access scheme consisting of both NOMA and OMA techniques. Specifically, NOMA is adopted for the two users within each user pair, while OMA is applied for different user pairs. In this subsection, three user pairing approaches are discussed to efficiently divide $2 N$ users into $N$ user pairs. 

\subsubsection{Channel-Based User Pairing}

Channel-based user pairing is the most popular user pairing approach in NOMA-based systems \cite{yin2016performance,almohimmah2018simple,janjua2020user}. The key to implement channel-based user pairing is to pair the two users which have more distinctive channel conditions. For the channel-based user pairing, the $2 N$ users are sorted based on their channel gains in the ascending order:
\begin{equation}
\setlength{\abovedisplayskip}{12pt}
\setlength{\belowdisplayskip}{12pt}
h_1^b \le \cdots \le h_k^b \le \cdots \le h_{2 N}^b,
\label{eqn:channel}
\end{equation}
where $h_k^b$ is given in (\ref{eqn:hk2}), and $b \in \{\textrm{d}, \textrm{u}\}$ with ``$\textrm{d}$'' and ``$\textrm{u}$'' denoting the downlink and uplink channels, respectively. After that, the sorted $2 N$ users are divided into two groups: the first group $G_1^{b, \textrm{c}}$ contains the first half of the sorted users starting from user 1 to user $N$; the second group $G_2^{b, \textrm{c}}$ consists of the second half starting from user $N + 1$ to user $2 N$. Hence, user pairing can be performed in the following manner: $U_i^{b, \textrm{c}} = \{G_1^{b, \textrm{c}}(i), G_2^{b, \textrm{c}}(i)\}$, i.e., the $i$-th user pair $U_i^{b, \textrm{c}}$ contains both the $i$-th user in $G_1^{b, \textrm{c}}$ and the $i$-th user in $G_2^{b, \textrm{c}}$ with $i = 1, 2, \cdots, N$.

Nevertheless, the channel-based user pairing approach only takes the channel conditions of different users into account, while the their specific QoS requirements are not considered.

\subsubsection{QoS-Based User Pairing}

In practical bidirectional LiFi-IoT systems, the supported users, i.e., IoT devices, might have their own distinctive QoS requirements in both downlink and uplink channels. Hence, the impact of users' QoS requirements should be considered when performing user pairing. For the QoS-based user pairing, all the $2 N$ users are sorted based on their QoS requirements in the descending order:
\begin{equation}
\setlength{\abovedisplayskip}{12pt}
\setlength{\belowdisplayskip}{12pt}
\widetilde{R}_1^b \ge \cdots \ge \widetilde{R}_k^b \ge \cdots \ge \widetilde{R}_{2 N}^b,
\label{eqn:QoS}
\end{equation}
where $\widetilde{R}_k^b$ denotes the QoS requirement, i.e., rate requirement, of the $k$-th user. Similarly, the first half and the second half of the sorted $2 N$ users can be divided into two groups which are denoted as $G_1^{b, \textrm{q}}$ and $G_2^{b, \textrm{q}}$, respectively. Hence, QoS-based user pairing can be given as follows: $U_i^{b, \textrm{q}} = \{G_1^{b, \textrm{q}}(i), G_2^{b, \textrm{q}}(i)\}$ with $i = 1, 2, \cdots, N$.

\subsubsection{Adaptive Channel and QoS-Based User Pairing}

\begin{algorithm}[!t]
    \caption{Adaptive channel and QoS-based user pairing}
    \label{alg:Allocation}
    \begin{algorithmic}[1]
        \STATE \textbf{Input:} $h_k^b$, $\widetilde{R}_k^b$, $P_z$, $b \in \{\textrm{d}, \textrm{u}\}$, $k = 1, 2, \cdots, 2 N$
        \STATE \textbf{Output:} optimal user pair $U_i^{b, \textrm{opt}}$, $i = 1, 2, \cdots, N$
        \STATE \textbf{Step 1: channel-based user pairing}
        \STATE Sort $\{h_k^b\}_{k = 1, 2, \cdots, 2 N}$ in ascending order
        \STATE Divide the sorted users into $G_1^{b, \textrm{c}}$ and $G_2^{b, \textrm{c}}$
        \STATE Obtain $U_i^{b, \textrm{c}} = \{G_1^{b, \textrm{c}}(i), G_2^{b, \textrm{c}}(i)\}$, $i = 1, 2, \cdots, N$
        \STATE Calculate $P_{\textrm{elec,min}}^{\textrm{c}}$ using $U_i^{b, \textrm{c}}$ and (\ref{eqn:Pmin})
        \STATE \textbf{Step 2: QoS-based user pairing}
        \STATE Sort $\{\widetilde{R}_k^b\}_{k = 1, 2, \cdots, 2 N}$ in descending order
        \STATE Divide the sorted users into $G_1^{b, \textrm{q}}$ and $G_2^{b, \textrm{q}}$
        \STATE Obtain $U_i^{b, \textrm{q}} = \{G_1^{b, \textrm{q}}(i), G_2^{b, \textrm{q}}(i)\}$, $i = 1, 2, \cdots, N$
        \STATE Calculate $P_{\textrm{elec,min}}^{\textrm{q}}$ using $U_i^{b, \textrm{q}}$ and (\ref{eqn:Pmin})
        \STATE \textbf{Step 3: adaptive selection}
        \FOR{$i = 1$ to $N$}
        \IF{$P_{\textrm{elec,min}}^{\textrm{c}} \le P_{\textrm{elec,min}}^{\textrm{q}}$}
        \STATE $U_i^{b, \textrm{opt}} = U_i^{b, \textrm{c}}$
        \ELSE
        \STATE $U_i^{b, \textrm{opt}} = U_i^{b, \textrm{q}}$
        \ENDIF
        \ENDFOR
    \end{algorithmic}
\end{algorithm}

In both the channel-based and the QoS-based user pairing approaches, only a single factor (i.e., channel gain or QoS requirement) is considered for all the users. However, due to the randomness of users' locations which leads to random channel gains and the randomness of users' QoS requirements, both channel gains and QoS requirements of users should be considered when putting them into pairs. 

In the following, an adaptive channel and QoS-based user pairing approach is proposed, which takes both users' channel gains and QoS requirements into consideration. The detailed procedures to perform adaptive channel and QoS-based user pairing are summarized in Algorithm~\ref{alg:Allocation}, which includes three steps. In the first step, channel-based user pairing is performed and the corresponding minimum power requirement $P_{\textrm{elec,min}}^{\textrm{c}}$ is calculated based on the obtained user pairs $U_i^{b, \textrm{c}}$ and (\ref{eqn:Pmin}). In the second step, QoS-based user pairing is conducted and the corresponding minimum power requirement $P_{\textrm{elec,min}}^{\textrm{q}}$ is obtained by utilizing $U_i^{b, \textrm{q}}$ and (\ref{eqn:Pmin}). In the third step, the user pairing approach which requires a lower minimum power is adaptively selected as the optimal one for the bidirectional LiFi-IoT system. Due to their randomness, users' locations and/or users' QoS requirements might change with time. The proposed adaptive user pairing approach is able to dynamically select the optimal one from the channel-based and the QoS-based user pairing approaches to achieve a lower minimum power consumption and hence a higher EE.

\section{Performance Analysis}

To fairly evaluate the performance of the bidirectional LiFi-IoT system using various multiple access techniques, we here adopt EE and UOP as two metrics for performance evaluation, which are analyzed in the following.

\subsection{Analysis of EE}

\subsubsection{EE Using NOMA with OPA}

For the bidirectional LiFi-IoT system using NOMA with OPA, the target sum rate and the required total electrical transmit power of all the $N$ pairs of users in both downlink and uplink channels are obtained by (\ref{eqn:R}) and (\ref{eqn:Pmin}), respectively. By substituting (\ref{eqn:R}) and (\ref{eqn:Pmin}) into (\ref{eqn:EE}), the EE of the bidirectional LiFi-IoT system using NOMA with OPA is given by
\begin{equation}
\setlength{\abovedisplayskip}{12pt}
\setlength{\belowdisplayskip}{12pt}
\eta_{\textrm{NOMA}}^{\textrm{OPA}} = \frac{\sum_{i = 1}^{N} \widetilde{R}_{i,f}^{\textrm{d}} + \widetilde{R}_{i,n}^{\textrm{d}} + \widetilde{R}_{i,f}^{\textrm{u}} + \widetilde{R}_{i,n}^{\textrm{u}}}{\sum_{i = 1}^{N} 2^{2 \widetilde{R}_{i,f}^{\textrm{d}}} \left( 2^{2 \widetilde{R}_{i,n}^{\textrm{d}}} \frac{P_z}{(h_{i,n}^{\textrm{d}})^2} + \frac{P_z}{(h_{i,f}^{\textrm{d}})^2 } \right) + 2^{2 \widetilde{R}_{i,n}^{\textrm{d}}} \frac{P_z}{(h_{i,n}^{\textrm{d}})^2} + 2^{2 \widetilde{R}_{i,f}^{\textrm{u}}} \frac{P_z}{(h_{i,f}^{\textrm{u}})^2} + 2^{2 \widetilde{R}_{i,n}^{\textrm{u}}} \frac{(1 + 2^{2 \widetilde{R}_{i,f}^{\textrm{u}}}) P_z}{(h_{i,n}^{\textrm{u}})^2}}. 
\label{eqn:EEQOMA}
\end{equation}

\subsubsection{EE Using NOMA with Channel-Based Power Allocation}

To compare the performance of NOMA with OPA and conventional NOMA with channel-based power allocation, the following two typical channel-based power allocation strategies are considered: (i) GRPA \cite{marshoud2016non} and (ii) NGDPA \cite{chen2018performance}. Letting $\alpha_i^{b}= \frac{p_{i,n}^{b,\textrm{NOMA}}}{p_{i,f}^{b,\textrm{NOMA}}}$ denote the power allocation ratio between the near user and the far user in the $i$-th user pair using NOMA with $b \in \{\textrm{d}, \textrm{u}\}$, the power allocation ratio using GRPA and NGDPA can be expressed by
\begin{equation}
\setlength{\abovedisplayskip}{12pt}
\setlength{\belowdisplayskip}{12pt}
\alpha_i^{b} = 
\left\{
\begin{aligned}
\left( \frac{h_{i,f}^{b}}{h_{i,n}^{b}} \right)^2,~~~~\textrm{GRPA}~~\\
\frac{h_{i,n}^{b} - h_{i,f}^{b}}{h_{i,n}^{b}},~~\textrm{NGDPA}\\
\end{aligned}~.
\right.
\label{eqn:GRPA}
\end{equation}

To satisfy the QoS requirements of both the far and near users in $i$-th user pair, following (\ref{eqn:pdu2}), the minimum required electrical transmit powers using NOMA with GRPA and NGDPA can be obtained by
\begin{equation}
\setlength{\abovedisplayskip}{12pt}
\setlength{\belowdisplayskip}{12pt}
  (p_{i,f}^{b,\textrm{NOMA}}, p_{i,n}^{b,\textrm{NOMA}}) = \left\{
  \begin{aligned}
   \left(p_{i,f}^{b,\textrm{opt}}, \alpha_i^{b} p_{i,f}^{b,\textrm{opt}}\right),~\alpha_i^{b} \ge \frac{p_{i,n}^{b,\textrm{opt}}}{p_{i,f}^{b,\textrm{opt}}}\\
    \left(\frac{p_{i,n}^{\textrm{d,opt}}}{\alpha_i^{b}}, p_{i,n}^{b,\textrm{opt}}\right),~ ~\alpha_i^{b} < \frac{p_{i,n}^{b,\textrm{opt}}}{p_{i,f}^{b,\textrm{opt}}}\\
  \end{aligned}~.
  \right.
\label{eqn:pduNOMA}
\end{equation}
Using (\ref{eqn:pduNOMA}), the minimum total electrical transmit power using NOMA with GRPA and NGDPA can be obtained. It can be clearly observed that the minimum required electrical transmit power using NOMA with GRPA and NGDPA is always larger or equal to that using NOMA with OPA. Therefore, the EE using NOMA with GRPA and NGDPA is always lower or equal to that using NOMA with OPA, which is omitted here for brevity.

\subsubsection{EE Using OMA}

For the bidirectional LiFi-IoT system using OMA techniques such as OFDMA and TDMA, as shown in Fig.~\ref{fig:Principle}, the QoS requirements of different users are satisfied by allocating them with different time or frequency resources. 

For the far and near users in the $i$-th user pair in both downlink and uplink channels using NOMA, the required rates to meet their QoS requirements are denoted by $\widetilde{R}_{i,f}^{b}$ and $\widetilde{R}_{i,n}^{b}$, respectively, with $b \in \{\textrm{d}, \textrm{u}\}$. When OMA is applied, to achieve the same individual rates as using NOMA, the required rates for both the far and near users in the $i$-th user pair is given by $\widetilde{R}_{i}^{b,\textrm{OMA}} = \widetilde{R}_{i,f}^{b} + \widetilde{R}_{i,n}^{b} $. Since the far and near users in the $i$-th user pair are not mutually interfered, their achievable rates can be obtained by
\begin{equation}
\setlength{\abovedisplayskip}{12pt}
\setlength{\belowdisplayskip}{12pt} 
  \left\{
  \begin{aligned}
  R_{i,f}^{b,\textrm{OMA}} = \frac{1}{2} \textrm{log}_2 \left( \frac{(h_{i,f}^{b})^2 p_{i,f}^{b,\textrm{OMA}}}{P_z} \right)~\\
  R_{i,n}^{b,\textrm{OMA}} = \frac{1}{2} \textrm{log}_2 \left( \frac{(h_{i,n}^{b})^2 p_{i,n}^{b,\textrm{OMA}}}{P_z} \right)~ \\
  \end{aligned}~,
  \right.
\label{eqn:snroma}
\end{equation}
where $p_{i,f}^{b,\textrm{OMA}}$ and $p_{i,n}^{b,\textrm{OMA}}$ denote the electrical transmit powers allocated to the far and near users in the $i$-th user pair, respectively. To ensure their QoS requirements, the following power requirements should be satisfied:
\begin{equation}
\setlength{\abovedisplayskip}{12pt}
\setlength{\belowdisplayskip}{12pt}
  \left\{
  \begin{aligned}
   p_{i,f}^{b,\textrm{OMA}} \ge 2^{2 (\widetilde{R}_{i,f}^{b} + \widetilde{R}_{i,n}^{b})} \frac{P_z}{(h_{i,f}^{b})^2 }\\
   p_{i,n}^{b,\textrm{OMA}} \ge 2^{2 (\widetilde{R}_{i,f}^{b} + \widetilde{R}_{i,n}^{b})} \frac{P_z}{(h_{i,n}^{b})^2} \\
  \end{aligned}~,
  \right.
\label{eqn:poma}
\end{equation}
and the minimum required electrical transmit powers for the far and near users in the $i$-th user pair are given by
\begin{equation}
\setlength{\abovedisplayskip}{12pt}
\setlength{\belowdisplayskip}{12pt}
  \left\{
  \begin{aligned}
   p_{i,f}^{b,\textrm{OMA,opt}} = 2^{2 (\widetilde{R}_{i,f}^{b} + \widetilde{R}_{i,n}^{b})} \frac{P_z}{(h_{i,f}^{b})^2 }\\
   p_{i,n}^{b,\textrm{OMA,opt}} = 2^{2 (\widetilde{R}_{i,f}^{b} + \widetilde{R}_{i,n}^{b})} \frac{P_z}{(h_{i,n}^{b})^2} \\
  \end{aligned}~.
  \right.
\label{eqn:poma2}
\end{equation}

Hence, the minimum total electrical transmit power using OMA is obtained by
\begin{equation}
\setlength{\abovedisplayskip}{12pt}
\setlength{\belowdisplayskip}{12pt}
P_{\textrm{elec,min}}^{\textrm{OMA}} = \sum_{i = 1}^{N} 2^{2 (\widetilde{R}_{i,f}^{\textrm{d}} + \widetilde{R}_{i,n}^{\textrm{d}})} \left( \frac{P_z}{(h_{i,f}^{\textrm{d}})^2 } + \frac{P_z}{(h_{i,n}^{\textrm{d}})^2 } \right) + 2^{2 (\widetilde{R}_{i,f}^{\textrm{u}} + \widetilde{R}_{i,n}^{\textrm{u}})} \left( \frac{P_z}{(h_{i,f}^{\textrm{u}})^2 } + \frac{P_z}{(h_{i,n}^{\textrm{u}})^2 } \right),
\label{eqn:PminOMA}
\end{equation}
and the EE of the bidirectional LiFi-IoT system using OMA can be obtained by
\begin{equation}
\setlength{\abovedisplayskip}{12pt}
\setlength{\belowdisplayskip}{12pt}
\eta_{\textrm{OMA}} = \frac{\sum_{i = 1}^{N} \widetilde{R}_{i,f}^{\textrm{d}} + \widetilde{R}_{i,n}^{\textrm{d}} + \widetilde{R}_{i,f}^{\textrm{u}} + \widetilde{R}_{i,n}^{\textrm{u}}}{\sum_{i = 1}^{N} 2^{2 (\widetilde{R}_{i,f}^{\textrm{d}} + \widetilde{R}_{i,n}^{\textrm{d}})} \left( \frac{P_z}{(h_{i,f}^{\textrm{d}})^2 } + \frac{P_z}{(h_{i,n}^{\textrm{d}})^2 } \right) + 2^{2 (\widetilde{R}_{i,f}^{\textrm{u}} + \widetilde{R}_{i,n}^{\textrm{u}})} \left( \frac{P_z}{(h_{i,f}^{\textrm{u}})^2 } + \frac{P_z}{(h_{i,n}^{\textrm{u}})^2 } \right)}. 
\label{eqn:EEOMA}
\end{equation}

In practical bidirectional LiFi-IoT systems, the minimum rate of each user can be achieved by using binary constellations such as binary phase shift keying (BPSK). As a result, considering the Hermitian symmetry constraint \cite{yin2016performance}, the practically minimum achievable rate of each user is given by $\frac{1}{2}$ bit/s/Hz. Based on the derived EEs of the bidirectional LiFi-IoT system using NOMA with OPA and OMA, we have the following proposition.

\begin{proposition}
The EE of the bidirectional LiFi-IoT system using NOMA with OPA is always larger or equal to that using OMA, i.e., $\eta_{\textrm{NOMA}}^{\textrm{OPA}} \ge \eta_{\textrm{OMA}}$, when $\widetilde{R}_{i,f}^{\textrm{d}}, \widetilde{R}_{i,n}^{\textrm{d}}, \widetilde{R}_{i,f}^{\textrm{u}}, \widetilde{R}_{i,n}^{\textrm{u}} \ge \frac{1}{2}$ (bit/s/Hz).
\end{proposition}

\begin{proof} Please refer to the appendix.
\end{proof}

Proposition 2 demonstrates that NOMA with OPA is more energy-efficient than OMA, which is very suitable for energy-sensitive IoT applications.

\subsection{Analysis of UOP}

Considering the maximum total downlink electrical transmit power constraint at the LiFi AP and the maximum uplink electrical transmit power constraint at each user, user outage might occur in both downlink and uplink channels of the bidirectional LiFi-IoT system. To evaluate the outage performance of the system, UOP is adopted as the metric in the following analysis.

For the downlink channel, the minimum required total electrical transmit power using NOMA with OPA is given by
\begin{equation}
\setlength{\abovedisplayskip}{12pt}
\setlength{\belowdisplayskip}{12pt}
P_{\textrm{elec,min}}^{\textrm{d}} = \sum_{i = 1}^{N} p_{i,f}^{\textrm{d,opt}} + p_{i,n}^{\textrm{d,opt}}.
\label{eqn:Pdmin}
\end{equation}
Due to the maximum total downlink electrical transmit power constraint at the LiFi AP, $P_{\textrm{elec,min}}^{\textrm{d}}$ cannot exceed its maximum value $P_{\textrm{max}}^{\textrm{d}}$. If $P_{\textrm{elec,min}}^{\textrm{d}}$ exceeds $P_{\textrm{max}}^{\textrm{d}}$, the LiFi AP will fail to support all the downlink users. Hence, the LiFi AP can only connect with a selected subset of the downlink users so as to meet the maximum total downlink electrical transmit power constraint. With the goal to let the LiFi AP connect with more downlink users, it is reasonable to exclude the users which require the highest powers outside the subset. 

 \begin{algorithm}[!t]
    \caption{Calculation of downlink UOP $P_{\textrm{out}}^{\textrm{d}}$}
    \label{alg:Poutd}
    \begin{algorithmic}[1]
        \STATE \textbf{Input:} $h_k^{\textrm{d}}$, $\widetilde{R}_k^{\textrm{d}}$, $P_z$, $P_{\textrm{max}}^{\textrm{d}}$, $k = 1, 2, \cdots, 2 N$
        \STATE \textbf{Output:} $P_{\textrm{out}}^{\textrm{d}}$
        \STATE Initialize $k_{\textrm{out}}^{\textrm{d}} = 0$
        \STATE Calculate $p_{i,f}^{\textrm{d,opt}}$ and $p_{i,n}^{\textrm{d,opt}}$ using (\ref{eqn:pdu2}), $i = 1, 2, \cdots, N$
        \STATE Sort $\{p_{i,f}^{\textrm{d,opt}}, p_{i,n}^{\textrm{d,opt}}\}_{i = 1, 2, \cdots, N}$ in descending order
        \STATE Obtain the sorted powers $\{p_k^{\textrm{d}}\}_{k = 1, 2, \cdots, 2 N}$
        \FOR{$k = 1$ to $2 N$}
        \STATE Calculate $P_k^{\textrm{d}} = \sum_{j = k}^{2 N} p_k^{\textrm{d}}$
        \IF{$P_k^{\textrm{d}} > P_{\textrm{max}}^{\textrm{d}}$}
        \STATE $k_{\textrm{out}}^{\textrm{d}} = k_{\textrm{out}}^{\textrm{d}} + 1$
        \ENDIF
        \ENDFOR 
        \STATE Calculate $P_{\textrm{out}}^{\textrm{d}} = \frac{k_{\textrm{out}}^{\textrm{d}}}{2 N}$
    \end{algorithmic}
\end{algorithm}

Let $k_{\textrm{out}}^{\textrm{d}}$ denote the number of downlink users that cannot connect with the LiFi AP, the downlink UOP can be calculated by $P_{\textrm{out}}^{\textrm{d}} = \frac{k_{\textrm{out}}^{\textrm{d}}}{2 N}$. The detailed procedures to calculate $P_{\textrm{out}}^{\textrm{d}}$ are given in Algorithm~\ref{alg:Poutd}. Due to the randomness of both users' locations and QoS requirements, $P_{\textrm{out}}^{\textrm{d}}$ is calculated for multiple times so as to obtain a stable average value.

For the uplink channel, the electrical transmit power of each user should not exceed the maximum value $p_{\textrm{max}}^{\textrm{u}}$. The calculation of the uplink UOP $P_{\textrm{out}}^{\textrm{u}}$ is hence straightforward. We only need to count the total number of uplink users which require a minimum electrical transmit power higher than $p_{\textrm{max}}^{\textrm{u}}$, i.e., $k_{\textrm{out}}^{\textrm{u}}$, and $P_{\textrm{out}}^{\textrm{u}} = \frac{k_{\textrm{out}}^{\textrm{u}}}{2 N}$. Similarly, $P_{\textrm{out}}^{\textrm{u}}$ is calculated  for multiple times to yield a stable average value.

Following the similar manner, both downlink and uplink UOPs of the bidirectional LiFi-IoT system using NOMA with channel-based power allocation and OMA can be also be achieved, which are omitted here for brevity.

\section{Performance Evaluation and Comparison} 

\subsection{Simulation Setup}

\begin{table}[!t]
\centering
\caption{SIMULATION PARAMETERS}\label{tab:Parameters}
\begin{tabular}{l|r}
\toprule
Parameter name, notation & Value\\
\midrule
Maximum cell radius, $r_{\mathrm{max}}$ & 3 m\\
Maximum vertical distance, $l_{\mathrm{max}}$ & 2.5 m\\
Minimum vertical distance, $l_{\mathrm{max}}$ & 1.5 m\\
Visible/infrared light LED semi-angle, $\Psi$ & $\textrm{70}^{\circ}$\\
Visible/infrared light PD responsivity, $\rho$ & 0.4 $\mathrm{A/W}$\\
Visible/infrared light PD active area, $A$ & 1 $\textrm{cm}^2$\\
Visible/infrared light PD FOV, $\Phi$ & $\textrm{70}^{\circ}$\\
Optical filter gain, $g_f$ & 0.9\\
Optical lens refractive index, $n$  & 1.5\\
Signal bandwidth, $B$ & 20 MHz\\
Noise PSD, $N_0$ & $10^{-22}~\mathrm{A^2/Hz}$\\
\bottomrule
\end{tabular}
\end{table}

In order to substantiate our derived analytical results, extensive Monte Carlo simulations are performed. If not otherwise specified, the simulation parameters of the considered single-cell bidirectional LiFi-IoT system are listed in Table I. For the purpose of performance comparison, four multiple access techniques are considered including: (i) OMA, (ii) NOMA with GRPA, (iii) NOMA with NGDPA and (iv) NOMA with OPA. Moreover, three user pairing approaches as discussed in Section III.E are evaluated which includes: (i) channel-based user pairing, (ii) QoS-based user pairing and (iii) adaptive channel and QoS-based user pairing. In the following evaluation, we assume that perfect channel state information (CSI) of all the users is available when performing user pairing and power allocation in the bidirectional LiFi-IoT system.

\subsection{Two-User Case}

\begin{figure}[!t]
\centering
{\includegraphics[width=0.9\columnwidth]{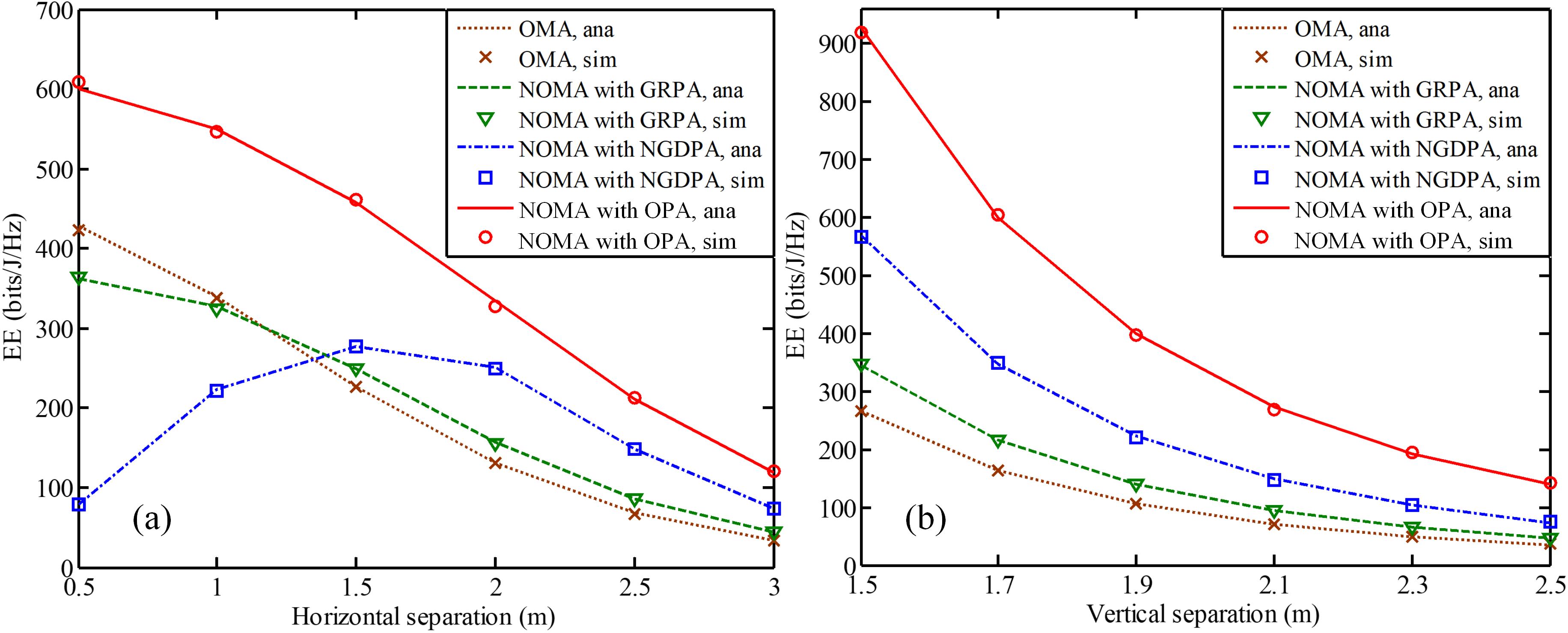}}
\caption{EE vs. (a) the horizontal separation and (b) the vertical separation of two users with $\widetilde{R}_{f}^{\textrm{d}}$ = $\widetilde{R}_{n}^{\textrm{d}}$ = $\widetilde{R}_{f}^{\textrm{u}}$ = $\widetilde{R}_{n}^{\textrm{u}}$ = 1 bit/s/Hz.} 
\label{fig:EE}
\end{figure}

We start with the case that there are only two users, i.e., a single pair of users, in the bidirectional LiFi-IoT system. Fig. \ref{fig:EE}(a) depicts the EE versus the horizontal separation of two users with $\widetilde{R}_{f}^{\textrm{d}}$ = $\widetilde{R}_{n}^{\textrm{d}}$ = $\widetilde{R}_{f}^{\textrm{u}}$ = $\widetilde{R}_{n}^{\textrm{u}}$ = 1 bit/s/Hz, where the near and far users have the same vertical distance $l_n$ = $l_f$ = 2.5 m, and the near user has a horizontal distance $r_n$ = 0 m while the horizontal distance of the far user $r_f$ varies from 0.5 to 3 m with a step of 0.5 m. For OMA, NOMA with GRPA and NOMA with OPA, the EE decreases with the increase of the horizontal separation. For NOMA with NGDPA, the EE first increases and then decreases with the increase of the horizontal separation, and the maximum EE of 276.3 bits/s/Hz is obtained at the horizontal separation of 1.5 m. It indicates that the performance of NOMA is largely dependent on the specific power allocation strategy adopted. Furthermore, NOMA with OPA can achieve the highest EE among all the four techniques. For example, the EE using NOMA with OPA is 458.1 bits/s/Hz at the horizontal separation of 1.5 m, which suggests an EE improvement of 65.8\% in comparison to that using NOMA with NGDPA. The EE versus the vertical separation of two users with $\widetilde{R}_{f}^{\textrm{d}}$ = $\widetilde{R}_{n}^{\textrm{d}}$ = $\widetilde{R}_{f}^{\textrm{u}}$ = $\widetilde{R}_{n}^{\textrm{u}}$ = 1 bit/s/Hz is shown in Fig. \ref{fig:EE}(b), where the near user has a vertical distance $l_n$ = 1.5 m and a horizontal distance $r_n$ = 0 m, while the vertical distance of the far user $l_f$  varies from 1.5 to 2.5 m with a step of 0.2 m and the corresponding horizontal distance of the far user is given by $r_f $ = 3$l_f$/2.5 so as to maintain a constant emission/incident angle. Fig. \ref{fig:EE}(b) demonstrates that the EE decreases with the increase of the vertical separation for all the four techniques and NOMA with OPA always achieves the highest EE. It can be observed from Figs. \ref{fig:EE}(a) and (b) that the simulation results are consistent with the analytical results.

\subsection{Multiple-User Case}

In the next, we further consider the multiple-user case, i.e., there are multiple pairs of users in the bidirectional LiFi-IoT system. Without loss of generality, we here assume that the vertical and horizontal distances of all the users are uniformly distributed between 1.5 to 2.5 m and 0 to 3 m, respectively. Moreover, we define $\widetilde{R}$ as the given QoS set for both downlink and uplink channels, where the elements of $\widetilde{R}$ have a unit of bits/s/Hz. Both the downlink and uplink QoS requirements of the users are randomly selected from the given QoS set $\widetilde{R}$. In order to obtain stable performance metrics, including EE and UOP, under random user locations and QoS requirements, we calculate the average EE and UOP over totally 10000 independent trials.

\begin{figure}[!t]
\centering
{\includegraphics[width=1\columnwidth]{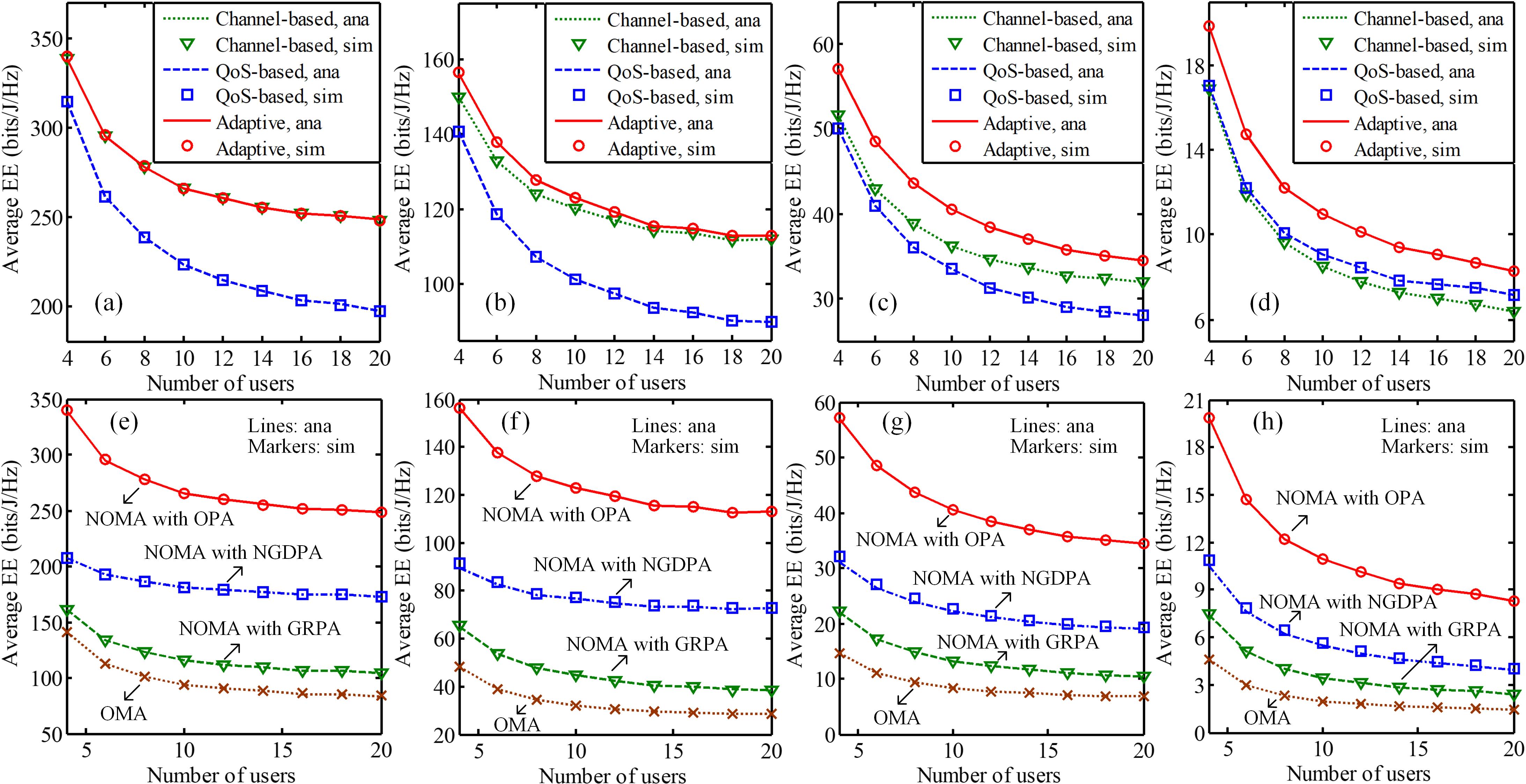}}
\caption{Average EE vs. the number of users using NOMA with OPA with different user pairing approaches for (a) $\widetilde{R} = 1$, (b) $\widetilde{R} = \{1, 2\}$, (c) $\widetilde{R} = \{1, 2, 3\}$ and (d) $\widetilde{R} = \{1, 2, 3, 4\}$, and average EE vs. the number of users using different multiple access techniques with adaptive channel and QoS-based user paring for (e) $\widetilde{R} = 1$, (f) $\widetilde{R} = \{1, 2\}$, (g) $\widetilde{R} = \{1, 2, 3\}$ and (h) $\widetilde{R} = \{1, 2, 3, 4\}$.} 
\label{fig:AEE}
\end{figure}

Figs. \ref{fig:AEE}(a)-(d) depict the average EE versus the number of users using NOMA with OPA with different user pairing approaches under various given QoS sets. When $\widetilde{R} = 1$, as shown in Fig. \ref{fig:AEE}(a), the average EE decreases with the increase of users for all the three user pairing approaches. Moreover, the channel-based user paring approach and the adaptive channel and QoS-based user paring approach obtain the same average EE, and both greatly outperform the QoS-based user paring approach. This is because the QoS requirements of all the user are the same, and hence random user pairing is achieved when sorting the users according to their QoS requirements. However, when diverse QoS requirements are desired by the users, the adaptive approach achieves higher average EE than the channel-based approach, as can be seen from Figs. \ref{fig:AEE}(b), (c) and (d). More specifically, when $\widetilde{R} = \{1, 2, 3, 4\}$, the QoS-based user paring approach outperforms the channel-based approach, especially for a relatively large number of users. It can be found from Figs. \ref{fig:AEE}(a)-(d) that it is beneficial to consider the impact of users' QoS requirements when performing user pairing and the proposed adaptive approach can be an efficient user pairing approach for energy-efficient bidirectional LiFi-IoT systems. Figs. \ref{fig:AEE}(e)-(h) compare the average EE performance of four different multiple access techniques utilizing adaptive channel and QoS-based user paring under various given QoS sets. As we can see, OMA always achieves the lowest average EE, while NOMA with GRPA slightly outperforms OMA. NOMA with NGDPA is shown to be more energy-saving than NOMA with GRPA, especially when users' QoS requirements are less diverse. Nevertheless, NOMA with OPA is proven to be the most energy-efficient one among all the four techniques, which can achieve a substantial average EE improvement in comparison to NOMA with NGDPA regardless of the number of users and the diversity of users' QoS requirements. It can also be seen from Fig. \ref{fig:AEE} that the simulation results agree well with the analytical results.

\begin{figure*}[!t]
\centering
{\includegraphics[width=1\columnwidth]{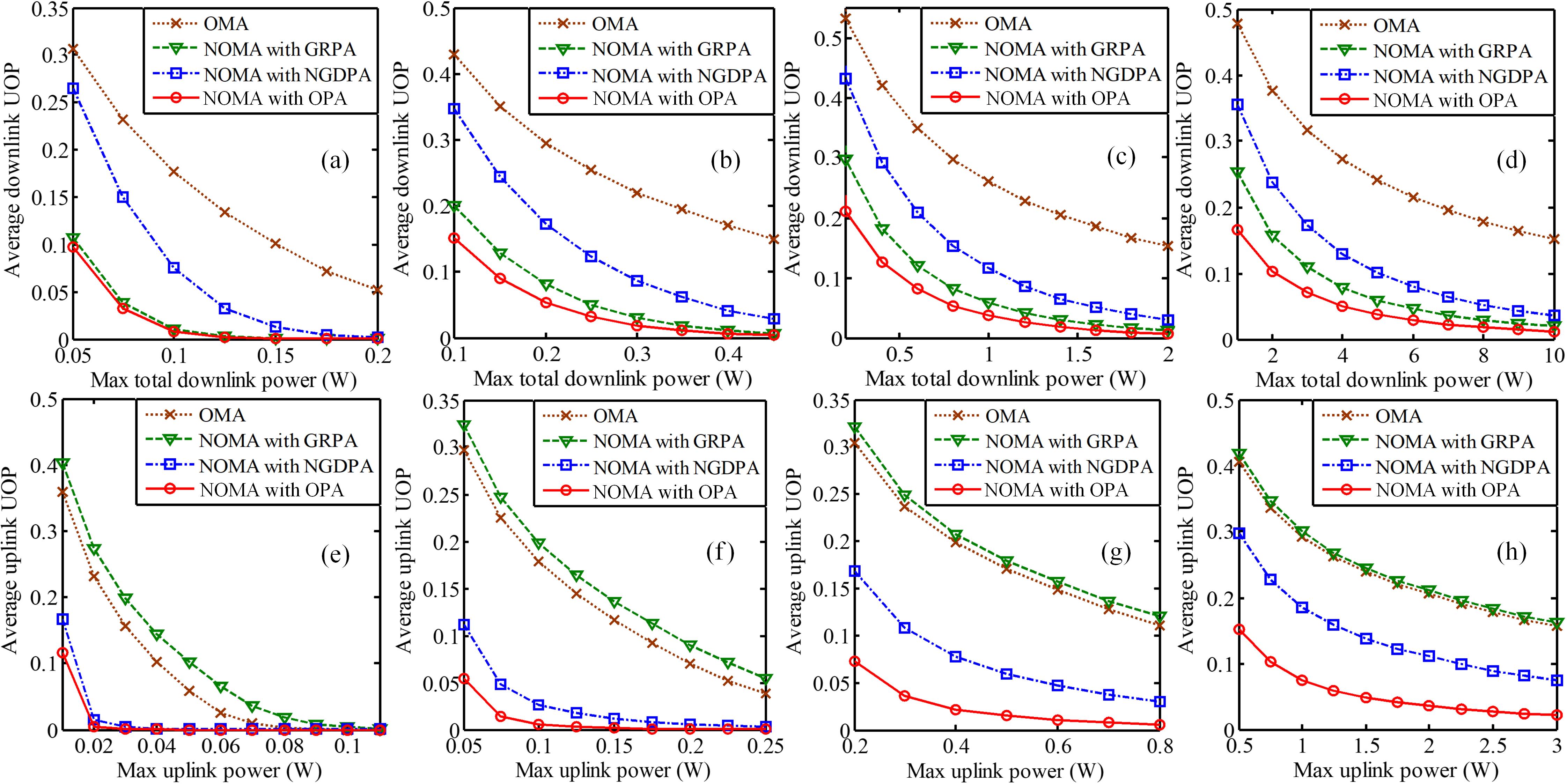}}
\caption{Average downlink UOP vs. the maximum total downlink power using different multiple access techniques with adaptive channel and QoS-based user paring for (a) $\widetilde{R} = 1$, (b) $\widetilde{R} = \{1, 2\}$, (c) $\widetilde{R} = \{1, 2, 3\}$ and (d) $\widetilde{R} = \{1, 2, 3, 4\}$, and average uplink UOP vs. the maximum uplink power using different multiple access techniques with adaptive channel and QoS-based user paring for (e) $\widetilde{R} = 1$, (f) $\widetilde{R} = \{1, 2\}$, (g) $\widetilde{R} = \{1, 2, 3\}$ and (h) $\widetilde{R} = \{1, 2, 3, 4\}$.} 
\label{fig:Pout2}
\end{figure*}

Figs. \ref{fig:Pout2}(a)-(d) show the average downlink UOP versus the maximum total downlink power using different multiple access techniques with adaptive channel and QoS-based user paring under various given QoS sets. It can be seen that OMA always obtains the highest average downlink UOP among all the four multiple access techniques. Moreover, NOMA with GRPA is shown to be more superior than NOMA with NGDPA in terms of average downlink UOP, especially for less diverse user QoS requirements and smaller maximum total downlink powers. Among them, NOMA with OPA achieves the lowest average downlink UOP regardless of the maximum total downlink power and the diversity of users' QoS requirements. Figs. \ref{fig:Pout2}(e)-(h) depict the average uplink UOP versus the maximum uplink power using NOMA with OPA using different user pairing approaches under various given QoS sets. It is interesting to see that NOMA with GRPA performs the worst among all the four multiple access techniques. In addition, OMA performs slightly better than NOMA with GRPA, while NOMA with NGDPA achieves comparable average uplink UOP as NOMA with OPA for $\widetilde{R} = 1$. When more diverse QoS requirements are desired by users, NOMA with OPA gradually outperforms NOMA with NGDPA and gives the lowest average uplink UOP. Fig. \ref{fig:Pout2} demonstrates that NOMA with OPA is more efficient than OMA and NOMA with GRPA/NGDPA to satisfactorily support multiple users in both downlink and uplink channels.

\section{Conclusion}
In this paper, we have proposed an energy-efficient NOMA technique for bidirectional LiFi-IoT communication, which adopts a QoS-guaranteed OPA strategy to maximize the system EE. We have identified and proved the optimal decoding orders in both downlink and uplink channels, and derived a closed-form OPA set. We have further proposed an adaptive channel and QoS-based user pairing approach which considers both users' channel gains and QoS requirements. The EE and UOP performance of the bidirectional LiFi-IoT system using different multiple access techniques have been analyzed. Extensive analytical and simulation results show that, compared with OMA and NOMA with channel-based power allocation and user pairing, NOMA adopting OPA with adaptive channel and QoS-based user pairing can significantly improve both the EE and UOP performance of the bidirectional LiFi-IoT system.

\section*{Appendix}

\subsection{\textit{Proof of Theorem 1}}

With the decoding order $\mathbb{O}_{i,\textrm{high}}^{\textrm{d}} \ge \mathbb{O}_{i,\textrm{low}}^{\textrm{d}}$, the high priority user decodes its message signal directly by treating the intended message signal for the low priority user as interference, while the low priority user needs to decode the intended message signal for the high priority user and apply SIC to decode its own message signal. Hence, the achievable rates of the high and low priority users in the $i$-th user pair can be given by
\begin{equation}
\setlength{\abovedisplayskip}{12pt}
\setlength{\belowdisplayskip}{12pt} 
  \left\{
  \begin{aligned}
  R_{i,\textrm{high}}^{\textrm{d}} = \frac{1}{2} \textrm{log}_2 \left( \frac{(h_{i,\textrm{high}}^{\textrm{d}})^2 p_{i,\textrm{high}}^{\textrm{d}}}{(h_{i,\textrm{high}}^{\textrm{d}})^2 p_{i,\textrm{low}}^{\textrm{d}} + P_z} \right) \\
  R_{i,\textrm{low}}^{\textrm{d}} = \frac{1}{2} \textrm{log}_2 \left( \frac{(h_{i,\textrm{low}}^{\textrm{d}})^2 p_{i,\textrm{low}}^{\textrm{d}}}{P_z} \right)~~~~~~~\\
  \end{aligned}~,
  \right.
\label{eqn:snrd}
\end{equation}
where the scaling factor $\frac{1}{2}$ is due to the Hermitian symmetry \cite{yin2016performance} and $P_z = N_0 B$ is the power of the additive noises. In addition, the achievable rate for the low priority user to decode the high priority user's message signal is obtained by
\begin{equation}
\setlength{\abovedisplayskip}{12pt}
\setlength{\belowdisplayskip}{12pt}
R_{{i,\textrm{low}} \rightarrow {\textrm{high}}}^{\textrm{d}} = \frac{1}{2} \textrm{log}_2 \left( \frac{(h_{i,\textrm{low}}^{\textrm{d}})^2 p_{i,\textrm{high}}^{\textrm{d}}}{(h_{i,\textrm{low}}^{\textrm{d}})^2 p_{i,\textrm{low}}^{\textrm{d}} + P_z} \right). 
\label{eqn:snrd2}
\end{equation}

To meet the QoS requirements, i.e., rate requirements, of both the high and low priority users in the $i$-th user pair, the following condition needs to be satisfied:
\begin{equation}
\setlength{\abovedisplayskip}{12pt}
\setlength{\belowdisplayskip}{12pt} 
\left\{
\begin{aligned}
\textrm{min} \{R_{i,\textrm{high}}^{\textrm{d}}, R_{{i,\textrm{low}} \rightarrow {i,\textrm{high}}}^{\textrm{d}}\} \ge \widetilde{R}_{i,\textrm{high}}^{\textrm{d}}\\
R_{i,\textrm{low}}^{\textrm{d}} \ge \widetilde{R}_{i,\textrm{low}}^{\textrm{d}}~~~~~~~~~~~~~~~~~~~~~~\\
\end{aligned}~.
\right.
\label{eqn:snrd3}
\end{equation}

According to (\ref{eqn:snrd3}), we can obtain the power requirements of two users as follows:
\begin{equation}
\setlength{\abovedisplayskip}{12pt}
\setlength{\belowdisplayskip}{12pt}
  \left\{
  \begin{aligned}
   p_{i,\textrm{high}}^{\textrm{d}} \ge \textrm{max} \left\{2^{2 \widetilde{R}_{i,\textrm{high}}^{\textrm{d}}} \left(p_{i,\textrm{low}}^{\textrm{d}} + \frac{P_z}{(h_{i,\textrm{high}}^{\textrm{d}})^2 } \right),  2^{2 \widetilde{R}_{i,\textrm{high}}^{\textrm{d}}} \left(p_{i,\textrm{low}}^{\textrm{d}} + \frac{P_z}{(h_{i,\textrm{low}}^{\textrm{d}})^2 } \right)\right\}\\
   p_{i,\textrm{low}}^{\textrm{d}} \ge 2^{2 \widetilde{R}_{i,\textrm{low}}^{\textrm{d}}} \frac{P_z}{(h_{i,\textrm{low}}^{\textrm{d}})^2}~~~~~~~~~~~~~~~~~~~~~~~~~~~~~~~~~~~~~~~~~~~~~~~~~~~~~~~~~~~\\
  \end{aligned}~.
  \right.
\label{eqn:pd2}
\end{equation}
Using $h_{i,\textrm{min}}^{\textrm{d}} = \textrm{min} \{h_{i,\textrm{high}}^{\textrm{d}}, h_{i,\textrm{low}}^{\textrm{d}}\}$, (\ref{eqn:pd2}) can be rewritten as
\begin{equation}
\setlength{\abovedisplayskip}{12pt}
\setlength{\belowdisplayskip}{12pt}
  \left\{
  \begin{aligned}
   p_{i,\textrm{high}}^{\textrm{d}} \ge 2^{2 \widetilde{R}_{i,\textrm{high}}^{\textrm{d}}} \left(p_{i,\textrm{low}}^{\textrm{d}} + \frac{P_z}{(h_{i,\textrm{min}}^{\textrm{d}})^2 } \right)\\
   p_{i,\textrm{low}}^{\textrm{d}} \ge 2^{2 \widetilde{R}_{i,\textrm{low}}^{\textrm{d}}} \frac{P_z}{(h_{i,\textrm{low}}^{\textrm{d}})^2}~~~~~~~~~~~~~~~\\
  \end{aligned}~.
  \right.
\label{eqn:pd3}
\end{equation}
By observing (\ref{eqn:pd3}), we can easily rewrite it into (\ref{eqn:pd}). Therefore, \textit{Theorem 1} is proved. $\hfill\blacksquare$ 

\subsection{\textit{Proof of Theorem 2}}

Based on the decoding order $\mathbb{O}_{i,\textrm{high}}^{\textrm{u}} \ge \mathbb{O}_{i,\textrm{low}}^{\textrm{u}}$, (\ref{eqn:yu2}) can be rewritten as:
\begin{equation}
\setlength{\abovedisplayskip}{12pt}
\setlength{\belowdisplayskip}{12pt}
y_i^{\textrm{u}} = h_{i,\textrm{high}}^{\textrm{u}} \sqrt{p_{i,\textrm{high}}^{\textrm{u}}} s_{i,\textrm{high}}^{\textrm{u}} + h_{i,\textrm{low}}^{\textrm{u}} \sqrt{p_{i,\textrm{low}}^{\textrm{u}}} s_{i,\textrm{low}}^{\textrm{u}} + z_{i}^{\textrm{u}}.
\label{eqn:yu3}
\end{equation}
Hence, the LiFi AP first decodes the high priority user's message signal directly by treating the low priority user's message signal, and then decodes the low priority user's message signal after applying SIC to remove the high priority user's message signal. Hence, the achievable rates of the high and low priority users in $i$-th user pair are obtained as follows:
\begin{equation}
\setlength{\abovedisplayskip}{12pt}
\setlength{\belowdisplayskip}{12pt} 
  \left\{
  \begin{aligned}
  R_{i,\textrm{high}}^{\textrm{u}} = \frac{1}{2} \textrm{log}_2 \left( \frac{(h_{i,\textrm{high}}^{\textrm{u}})^2 p_{i,\textrm{high}}^{\textrm{u}}}{(h_{i,\textrm{low}}^{\textrm{u}})^2 p_{i,\textrm{low}}^{\textrm{u}} + P_z} \right)\\
  R_{i,\textrm{low}}^{\textrm{u}} = \frac{1}{2} \textrm{log}_2 \left( \frac{(h_{i,\textrm{low}}^{\textrm{u}})^2 p_{i,\textrm{low}}^{\textrm{u}}}{P_z} \right)~~~~~~~\\
  \end{aligned}~.
  \right.
\label{eqn:snru}
\end{equation}

To meet the QoS requirements for the LiFi AP to successfully decode the intended message signals for both the high and low priority users in $i$-th user pair, the following condition needs to be satisfied:
\begin{equation}
\setlength{\abovedisplayskip}{12pt}
\setlength{\belowdisplayskip}{12pt} 
  \left\{
  \begin{aligned}
  R_{i,\textrm{high}}^{\textrm{u}} \ge \widetilde{R}_{i,\textrm{high}}^{\textrm{u}} \\
  R_{i,\textrm{low}}^{\textrm{u}} \ge \widetilde{R}_{i,\textrm{low}}^{\textrm{u}} ~\\
  \end{aligned}~.
  \right.
\label{eqn:snru2}
\end{equation}

According to (\ref{eqn:snru2}), we can have the power requirement of the high and low priority users in $i$-th user pair:
\begin{equation}
\setlength{\abovedisplayskip}{12pt}
\setlength{\belowdisplayskip}{12pt}
  \left\{
  \begin{aligned}
   p_{i,\textrm{high}}^{\textrm{u}} \ge 2^{2 \widetilde{R}_{i,\textrm{high}}^{\textrm{u}}} \frac{(h_{i,\textrm{low}}^{\textrm{u}})^2 p_{i,\textrm{low}}^{\textrm{u}} + P_z}{(h_{i,\textrm{high}}^{\textrm{u}})^2}\\
   p_{i,\textrm{low}}^{\textrm{u}} \ge 2^{2 \widetilde{R}_{i,\textrm{low}}^{\textrm{u}}} \frac{P_z}{(h_{i,\textrm{low}}^{\textrm{u}})^2}~~~~~~~~~~~~~\\
  \end{aligned}~.
  \right.
\label{eqn:pu2}
\end{equation}
Substituting $(h_{i,\textrm{low}}^{\textrm{u}})^2 p_{i,\textrm{low}}^{\textrm{u}} \ge 2^{2 \widetilde{R}_{i,\textrm{low}}^{\textrm{u}}} P_z$ into (\ref{eqn:pu2}) yields (\ref{eqn:pu}). Hence, the proof of \textit{Theorem 2} is completed. $\hfill\blacksquare$ 

\subsection{\textit{Proof of Proposition 1}}

Since there are only two users in the $i$-th user pair, the decoding orders for both downlink and uplink channels can only have two options: i.e., $\mathbb{O}_{i,f}^{\textrm{d}} \ge \mathbb{O}_{i,n}^{\textrm{d}}$ and $\mathbb{O}_{i,f}^{\textrm{d}} < \mathbb{O}_{i,n}^{\textrm{d}}$ for downlink, and $\mathbb{O}_{i,f}^{\textrm{u}} \ge \mathbb{O}_{i,n}^{\textrm{u}}$ and $\mathbb{O}_{i,f}^{\textrm{u}} < \mathbb{O}_{i,n}^{\textrm{u}}$ for uplink.

For the downlink with $\mathbb{O}_{i,f}^{\textrm{d}} \ge \mathbb{O}_{i,n}^{\textrm{d}}$, using (\ref{eqn:pd}), the power requirements can be obtained by
\begin{equation}
\setlength{\abovedisplayskip}{12pt}
\setlength{\belowdisplayskip}{12pt}
  \left\{
  \begin{aligned}
   p_{i,f}^{\textrm{d}} \ge 2^{2 \widetilde{R}_{i,f}^{\textrm{d}}} \left( 2^{2 \widetilde{R}_{i,n}^{\textrm{d}}} \frac{P_z}{(h_{i,n}^{\textrm{d}})^2} + \frac{P_z}{(h_{i,\textrm{min}}^{\textrm{d}})^2 } \right)\\
   p_{i,n}^{\textrm{d}} \ge 2^{2 \widetilde{R}_{i,n}^{\textrm{d}}} \frac{P_z}{(h_{i,n}^{\textrm{d}})^2}~~~~~~~~~~~~~~~~~~~~~~~~~\\
  \end{aligned}~,
  \right.
\label{eqn:pdo1}
\end{equation}
where $h_{i,\textrm{min}}^{\textrm{d}} = h_{i,f}^{\textrm{d}}$. Hence, the total required power of both the far and near users in the $i$-th user pair in the downlink with $\mathbb{O}_{i,f}^{\textrm{d}} \ge \mathbb{O}_{i,n}^{\textrm{d}}$ is obtained by
\begin{equation}
\setlength{\abovedisplayskip}{12pt}
\setlength{\belowdisplayskip}{12pt}
p_{i}^{\textrm{d}} =  p_{i,f}^{\textrm{d}} +  p_{i,n}^{\textrm{d}} \ge 2^{2 (\widetilde{R}_{i,f}^{\textrm{d}} + \widetilde{R}_{i,n}^{\textrm{d}})} \frac{P_z}{(h_{i,n}^{\textrm{d}})^2} + 2^{2 \widetilde{R}_{i,f}^{\textrm{d}}} \frac{P_z}{(h_{i,f}^{\textrm{d}})^2 } + 2^{2 \widetilde{R}_{i,n}^{\textrm{d}}} \frac{P_z}{(h_{i,n}^{\textrm{d}})^2}.
\label{eqn:pdtot1}
\end{equation}

Similarly, the total required power of both the far and near users in the $i$-th user pair in the downlink with $\mathbb{O}_{i,f}^{\textrm{d}} < \mathbb{O}_{i,n}^{\textrm{d}}$ can also be achieved by
\begin{equation}
\setlength{\abovedisplayskip}{12pt}
\setlength{\belowdisplayskip}{12pt}
p_{i}^{\textrm{d},\dagger} \ge 2^{2 (\widetilde{R}_{i,f}^{\textrm{d}} + \widetilde{R}_{i,n}^{\textrm{d}})} \frac{P_z}{(h_{i,f}^{\textrm{d}})^2} + 2^{2 \widetilde{R}_{i,f}^{\textrm{d}}} \frac{P_z}{(h_{i,f}^{\textrm{d}})^2 } + 2^{2 \widetilde{R}_{i,n}^{\textrm{d}}} \frac{P_z}{(h_{i,n}^{\textrm{d}})^2}.
\label{eqn:pdtot2}
\end{equation}
Since $h_{i,f}^{\textrm{d}} \le h_{i,n}^{\textrm{d}}$, by observing (\ref{eqn:pdtot1}) and (\ref{eqn:pdtot2}), we can find that the minimum power requirement with $\mathbb{O}_{i,f}^{\textrm{d}} < \mathbb{O}_{i,n}^{\textrm{d}}$ is larger or equal to that with $\mathbb{O}_{i,f}^{\textrm{d}} \ge \mathbb{O}_{i,n}^{\textrm{d}}$. As a result, $\mathbb{O}_{i,f}^{\textrm{d}} \ge \mathbb{O}_{i,n}^{\textrm{d}}$ is the optimal decoding order for the downlink.

Next, for the uplink with $\mathbb{O}_{i,f}^{\textrm{u}} \ge \mathbb{O}_{i,n}^{\textrm{u}}$, using (\ref{eqn:pu}), the power requirements can be achieved by
\begin{equation}
\setlength{\abovedisplayskip}{12pt}
\setlength{\belowdisplayskip}{12pt}
  \left\{
  \begin{aligned}
   p_{i,f}^{\textrm{u}} \ge 2^{2 \widetilde{R}_{i,f}^{\textrm{u}}} \frac{(1 + 2^{2 \widetilde{R}_{i,n}^{\textrm{u}}}) P_z}{(h_{i,f}^{\textrm{u}})^2}\\
   p_{i,n}^{\textrm{u}} \ge 2^{2 \widetilde{R}_{i,n}^{\textrm{u}}} \frac{P_z}{(h_{i,n}^{\textrm{u}})^2}~~~~~~~~~\\
  \end{aligned}~,
  \right.
\label{eqn:puo1}
\end{equation}
and the total required power of both the far and near users in the $i$-th user pair in the uplink with $\mathbb{O}_{i,f}^{\textrm{u}} \ge \mathbb{O}_{i,n}^{\textrm{u}}$ is given by
\begin{equation}
\setlength{\abovedisplayskip}{12pt}
\setlength{\belowdisplayskip}{12pt}
p_{i}^{\textrm{u}} =  p_{i,f}^{\textrm{u}} +  p_{i,n}^{\textrm{u}} \ge 2^{2 (\widetilde{R}_{i,f}^{\textrm{u}} + \widetilde{R}_{i,n}^{\textrm{u}})} \frac{P_z}{(h_{i,f}^{\textrm{u}})^2} + 2^{2 \widetilde{R}_{i,f}^{\textrm{u}}} \frac{P_z}{(h_{i,f}^{\textrm{u}})^2} + 2^{2 \widetilde{R}_{i,n}^{\textrm{u}}} \frac{P_z}{(h_{i,n}^{\textrm{u}})^2}.
\label{eqn:putot1}
\end{equation}
Similarly, we can also obtain the total power requirement of both the far and near users in the $i$-th user pair in the uplink with $\mathbb{O}_{i,f}^{\textrm{u}} < \mathbb{O}_{i,n}^{\textrm{u}}$ as follows:
\begin{equation}
\setlength{\abovedisplayskip}{12pt}
\setlength{\belowdisplayskip}{12pt}
p_{i}^{\textrm{u},\dagger} \ge 2^{2 (\widetilde{R}_{i,f}^{\textrm{u}} + \widetilde{R}_{i,n}^{\textrm{u}})} \frac{P_z}{(h_{i,n}^{\textrm{u}})^2} + 2^{2 \widetilde{R}_{i,f}^{\textrm{u}}} \frac{P_z}{(h_{i,f}^{\textrm{u}})^2} + 2^{2 \widetilde{R}_{i,n}^{\textrm{u}}} \frac{P_z}{(h_{i,n}^{\textrm{u}})^2}.
\label{eqn:putot2}
\end{equation}
Using $h_{i,f}^{\textrm{u}} \le h_{i,n}^{\textrm{u}}$ and observing (\ref{eqn:putot1}) and (\ref{eqn:putot2}), it can be seen that the minimum power requirement with $\mathbb{O}_{i,f}^{\textrm{u}} \ge \mathbb{O}_{i,n}^{\textrm{u}}$ is larger or equal to that with $\mathbb{O}_{i,f}^{\textrm{u}} < \mathbb{O}_{i,n}^{\textrm{u}}$. Hence, $\mathbb{O}_{i,f}^{\textrm{u}} < \mathbb{O}_{i,n}^{\textrm{u}}$ is the optimal decoding order for the uplink. Therefore, \textit{Proposition 1} is proved. $\hfill\blacksquare$ 

\subsection{\textit{Proof of Proposition 2}}

By observing (\ref{eqn:EEQOMA}) and (\ref{eqn:EEOMA}), to prove that $\eta_{\textrm{NOMA}}^{\textrm{OPA}} \ge \eta_{\textrm{OMA}}$, we only need to prove $P_{\textrm{elec,min}}^{\textrm{NOMA}} \le P_{\textrm{elec,min}}^{\textrm{OMA}}$. Using (\ref{eqn:Pmin}) and (\ref{eqn:PminOMA}), the difference between $P_{\textrm{elec,min}}^{\textrm{NOMA}}$ and $P_{\textrm{elec,min}}^{\textrm{OMA}}$ is given by
\begin{equation}
\setlength{\abovedisplayskip}{12pt}
\setlength{\belowdisplayskip}{12pt}
\begin{split}
P_{\textrm{elec,min}}^{\textrm{OMA}}- P_{\textrm{elec,min}}^{\textrm{NOMA}}
& = \sum_{i = 1}^{N} 2^{2 (\widetilde{R}_{i,f}^{\textrm{d}} + \widetilde{R}_{i,n}^{\textrm{d}})} \frac{P_z}{(h_{i,f}^{\textrm{d}})^2} - 2^{2 \widetilde{R}_{i,f}^{\textrm{d}}} \frac{P_z}{(h_{i,f}^{\textrm{d}})^2} - 2^{2 \widetilde{R}_{i,n}^{\textrm{d}}} \frac{P_z}{(h_{i,n}^{\textrm{d}})^2} \\
&~~~~~+ 2^{2 (\widetilde{R}_{i,f}^{\textrm{u}} + \widetilde{R}_{i,n}^{\textrm{u}})} \frac{P_z}{(h_{i,f}^{\textrm{u}})^2} - 2^{2 \widetilde{R}_{i,f}^{\textrm{u}}} \frac{P_z}{(h_{i,f}^{\textrm{u}})^2} - 2^{2 \widetilde{R}_{i,n}^{\textrm{u}}} \frac{P_z}{(h_{i,n}^{\textrm{u}})^2}.
\end{split}
\label{eqn:Pdiff}
\end{equation}
Since $h_{i,f}^{\textrm{d}} \le h_{i,n}^{\textrm{d}}$ and $h_{i,f}^{\textrm{u}} \le h_{i,n}^{\textrm{u}}$, we have
\begin{equation}
\setlength{\abovedisplayskip}{12pt}
\setlength{\belowdisplayskip}{12pt}
\begin{split}
P_{\textrm{elec,min}}^{\textrm{OMA}}- P_{\textrm{elec,min}}^{\textrm{NOMA}}
& \ge \sum_{i = 1}^{N} 2^{2 (\widetilde{R}_{i,f}^{\textrm{d}} + \widetilde{R}_{i,n}^{\textrm{d}})} \frac{P_z}{(h_{i,f}^{\textrm{d}})^2} - 2^{2 \widetilde{R}_{i,f}^{\textrm{d}}} \frac{P_z}{(h_{i,f}^{\textrm{d}})^2} - 2^{2 \widetilde{R}_{i,n}^{\textrm{d}}} \frac{P_z}{(h_{i,f}^{\textrm{d}})^2} \\
&~~~~~+ 2^{2 (\widetilde{R}_{i,f}^{\textrm{u}} + \widetilde{R}_{i,n}^{\textrm{u}})} \frac{P_z}{(h_{i,f}^{\textrm{u}})^2} - 2^{2 \widetilde{R}_{i,f}^{\textrm{u}}} \frac{P_z}{(h_{i,f}^{\textrm{u}})^2} - 2^{2 \widetilde{R}_{i,n}^{\textrm{u}}} \frac{P_z}{(h_{i,f}^{\textrm{u}})^2}.
\end{split}
\label{eqn:Pdiff2}
\end{equation}
Let $\widetilde{R}_{i,\textrm{max}}^{b} = \textrm{max} \{\widetilde{R}_{i,f}^{b}, \widetilde{R}_{i,n}^{b}\} $ and $\widetilde{R}_{i,\textrm{min}}^{b} = \textrm{min} \{\widetilde{R}_{i,f}^{b}, \widetilde{R}_{i,n}^{b}\} $ with $b \in \{\textrm{d}, \textrm{u}\}$, and hence (\ref{eqn:Pdiff2}) can be written as:
\begin{equation}
\setlength{\abovedisplayskip}{12pt}
\setlength{\belowdisplayskip}{12pt}
\begin{split}
P_{\textrm{elec,min}}^{\textrm{OMA}}- P_{\textrm{elec,min}}^{\textrm{NOMA}}
& \ge \sum_{i = 1}^{N} \frac{P_z}{(h_{i,f}^{\textrm{d}})^2} \left( 2^{2 \widetilde{R}_{i,\textrm{max}}^{\textrm{d}}} \left(2^{2 \widetilde{R}_{i,\textrm{min}}^{\textrm{d}}} - 1 \right) - 2^{2 \widetilde{R}_{i,\textrm{min}}^{\textrm{d}}}\right)\\
&~~~~~+ \frac{P_z}{(h_{i,f}^{\textrm{u}})^2} \left( 2^{2 \widetilde{R}_{i,\textrm{max}}^{\textrm{u}}} \left(2^{2 \widetilde{R}_{i,\textrm{min}}^{\textrm{u}}} - 1 \right) - 2^{2 \widetilde{R}_{i,\textrm{min}}^{\textrm{u}}}\right).
\end{split}
\label{eqn:Pdiff3}
\end{equation}
Since $2^{2 \widetilde{R}_{i,\textrm{min}}^{b}} - 1 \ge 1$ due to $\widetilde{R}_{i,\textrm{min}}^{b} \ge \frac{1}{2}$, we have
\begin{equation}
\setlength{\abovedisplayskip}{12pt}
\setlength{\belowdisplayskip}{12pt}
P_{\textrm{elec,min}}^{\textrm{OMA}}- P_{\textrm{elec,min}}^{\textrm{NOMA}} \ge \sum_{i = 1}^{N} \frac{P_z}{(h_{i,f}^{\textrm{d}})^2} \left( 2^{2 \widetilde{R}_{i,\textrm{max}}^{\textrm{d}}} - 2^{2 \widetilde{R}_{i,\textrm{min}}^{\textrm{d}}}\right) + \frac{P_z}{(h_{i,f}^{\textrm{u}})^2} \left( 2^{2 \widetilde{R}_{i,\textrm{max}}^{\textrm{u}}} - 2^{2 \widetilde{R}_{i,\textrm{min}}^{\textrm{u}}}\right).
\label{eqn:Pdiff4}
\end{equation}
Since $\widetilde{R}_{i,\textrm{max}}^{b} \ge \widetilde{R}_{i,\textrm{min}}^{b}$, we have $2^{2 \widetilde{R}_{i,\textrm{max}}^{b}} - 2^{2 \widetilde{R}_{i,\textrm{min}}^{b}} \ge 0$, and hence we obtain $P_{\textrm{elec,min}}^{\textrm{OMA}}- P_{\textrm{elec,min}}^{\textrm{NOMA}} \ge 0 $. Therefore, the proof of \textit{Proposition 2} is completed. $\hfill\blacksquare$

\bibliographystyle{IEEEtran}
\bibliography{IEEEabrv,mylib}

\end{document}